\documentclass{elsart}
\usepackage{times}
\usepackage{epsfig}
\usepackage{marvosym}
\usepackage{amssymb}
\begin{document}
\bibliographystyle{roman}

\journal{Nuclear Instruments and Methods A}

%%%%%%%%%%%%%%%%%%%%%%%%%%%%%%%%%%%%%%%%%%%%%%%%%%%%%%%%%%%%%%%%%%%%%%%%%%%%%%%
\def\hb{\hfill\break}
\def\MeV{\rm MeV}
\def\GeV{\rm GeV}
\def\TeV{\rm TeV}

\def\m{\rm m}
\def\cm{\rm cm}
\def\mm{\rm mm}
\def\lam{$\lambda_{\rm int}$}
\def\rad{$X_0$}
 
\def\NIM{Nucl. Instr. and Meth.~}
\def\ieee {{IEEE Trans. Nucl. Sci.~}}
\def\prl{Phys. Rev. Lett.~}

\def\etal{{\it et al.}}
\def\eg{{\it e.g.,~}}
\def\ie{{\it i.e.,~}}
\def\cf{{\it cf.~}}
\def\etc{{\it etc.~}}
\def\vs{{\it vs.~}}

\hyphenation{ca-lo-ri-me-ter}
\hyphenation{ca-lo-ri-me-ters}
\hyphenation{Brems-strah-lung}

%%%%%%%%%%%%%%%%%%%%%%%%%%%%%%%%%%%%%%%%%%%%%%%%%%%%%%%%%%%%%%%%%%%%%%%%%%%%%%%
\begin{frontmatter}
\title{Hadron detection with a dual-readout fiber calorimeter}

\author{S. Lee$^m$, A. Cardini$^c$, M. Cascella$^d$, S. Choi$^e$, G. Ciapetti$\dagger$$^g$,} 
\author{R. Ferrari$^h$, S. Franchino$^i$, M.~Fraternali$^f$, G.~Gaudio$^h$, S.~Ha$^e$,} 
\author{J. Hauptman$^j$, H. Kim$^m$, A.~Lanza$^h$, F.~Li$^j$, M. Livan$^f$, E. Meoni$^{l}$,}
\author{J.~Park$^m$, F. Scuri$^b$, A. Sill$^a$ and R. Wigmans$^{a,}$\thanksref{Corres}} 

\address{$^a$ Texas Tech University, Lubbock (TX), USA\\
$^b$ INFN Sezione di Pisa, Italy\\
$^c$ INFN Sezione di Cagliari, Monserrato (CA), Italy\\
$^d$ University College, London, UK\\
$^e$ Korea University, Seoul, Korea\\
$^f$ INFN Sezione di Pavia and Dipartimento di Fisica, Universit\`a di Pavia, Italy\\
$^g$ Dipartimento di Fisica, Universit\`a di Roma ''La Sapienza''  and INFN Sezione di Roma, Italy\\
$^h$ INFN Sezione di Pavia, Italy\\
$^i$ Kirchhoff-Institut f\"ur Physik, Ruprecht-Karls-Universit\"at Heidelberg, Germany\\
$^j$ Iowa State University, Ames (IA), USA\\
$^l$ Dipartimento di Fisica, Universit\`a della Calabria and INFN Cosenza, Italy\\
$^m$ Kyungpook National University, Daegu, Korea\\
$^\dagger$ Deceased}
%$^l$ CERN, Gen\`eve, Switzerland

%\thanks[Leave1]{Now at Department of Physics and INFN, University of Milano Statale, Italy}
\thanks[Corres]{Corresponding author.
              Email wigmans@ttu.edu, fax (+1) 806 742-1182.}
              
%%%%%%%%%%%%%%%%%%%%%%%%%%%
%%%%%%%%%%%%%%%%%%%%%%%%%%%
\vskip -7mm
\begin{abstract}
In this paper, we describe measurements of the response functions of a fiber-based dual-readout 
calorimeter for pions, protons and multiparticle ``jets'' with energies in the range from 10 to 180 GeV.
The calorimeter uses lead as absorber material and has a total mass of 1350 kg. It is complemented by
leakage counters made of scintillating plastic, with a total mass of 500 kg. The effects of these leakage counters 
on the calorimeter performance are studied as well. In a separate section, we investigate and compare different methods
to measure the energy resolution of a calorimeter. Using only the signals provided by the calorimeter, we demonstrate
that our dual-readout calorimeter, calibrated with electrons,  is able to reconstruct the energy of proton and pion beam
particles to within a few percent at all energies. The fractional widths of the signal distributions for these particles ($\sigma/E$) scale with the beam energy as $30\%/\sqrt{E}$, without any additional contributing terms.

\vskip 3mm
\noindent
{\it PACS:} 29.40.Ka, 29.40.Mc, 29.40.Vj
\vskip -5mm
\end{abstract}
\begin{keyword}
Dual-readout calorimetry, \v{C}erenkov light, optical fibers
\end{keyword}
\end{frontmatter}
%\vskip -8mm
%{\sl * This paper is dedicated to the memory of our long-time friend and collaborator Guido Ciapetti}
\newpage
%%%%%%%%%%%%%%%%%%%%%%%%%%%%%%%%%%%%%%%%%%%%%%%%%%%%%%%%%%%%%%%%%%%%%%%%%%%%%%%
%\linenumbers
\section{Introduction}
\vskip -5mm

The performance of hadron calorimeters is typically strongly dominated, and negatively affected, by the effects of fluctuations in the electromagnetic (em) shower fraction, $f_{\rm em}$.
One approach to eliminate the effects of such fluctuations is to {\sl measure} $f_{\rm em}$ for each event. It turns out that the \v{C}erenkov mechanism provides unique opportunities in this respect.
Calorimeters that use \v{C}erenkov light as signal source are, for all practical purposes, only
responding to the em fraction of hadronic showers \cite{Akc97}. 
%This is because the electrons/positrons through which the energy is deposited in the em shower component are relativistic down to energies
%of only 200 keV. On the other hand, most of the non-em energy in hadron showers is deposited by
%non-relativistic protons generated in nuclear reactions.
%Such protons do generate signals in active media such as plastic scintillator or liquid argon. 
By comparing the relative strengths of the 
signals representing the visible deposited energy and the \v{C}erenkov light produced in the shower absorption process, the em shower fraction can be determined and the total shower energy can be reconstructed using the known $e/h$ value(s) of the calorimeter\footnote{The ratio $e/h$ represents the ratio of the average calorimeter signals per unit deposited energy from the em and non-em components of hadron showers.
A calorimeter with $e/h = 1$ is said to be {\sl compensating}, but in practice almost all calorimeters have $e/h > 1$.}. This is the essence of what has become known as {\sl dual-readout }calorimetry. We are studying the properties of particle detectors of this type in the context of CERN's RD52 project \cite{RD52web}.

In the dual-readout calorimeter discussed in this paper, signals are generated in scintillating fibers, which measure the deposited energy, and in clear plastic fibers, which measure the relativistic shower particles, by means of the \v{C}erenkov light generated by these. A large number of such fibers are embedded in a metal absorber structure.
This detector is longitudinally unsegmented, the fibers are oriented in {\sl approximately} the same direction as the particles to be detected. 
In previous papers, we have focused on the electromagnetic performance of such a detector \cite{RD52_em,smalltheta} and on its capability to identify the particles developing showers in it \cite{PID}. 

In this paper, we describe experiments in which the hadronic performance of this calorimeter was measured. Hadron showers require a very large volume to fully develop. The 1350 kg fiducial volume of the calorimeter used for our purpose absorbed in practice, on average, only $\sim 90\%$ of the shower, depending on the energy of the showering particle. Therefore, fluctuations in (side) leakage formed a dominating contribution to the energy resolution. In order to get a handle on this contribution, the calorimeter was surrounded by a (rather crude) system of leakage counters. In our measurements, we also tried to distinguish between showers initiated by pions and by protons, using the calorimeter information. Our experimental program concentrated on two issues:
\begin{enumerate}
\item To what extent can the very crude system of leakage counters that we had installed around the calorimeter measure these event-to-event fluctuations and improve the
measured energy resolution?
\item Can we separate pions and protons in the CERN SPS H8 beam, and measure the dual-readout calorimeter performance separately for these particles?
\end{enumerate}
We also studied the performance for multi-particle ``jets,'' produced in high-multi-plicity interactions by the beam hadrons in an upstream target. In modern particle physics experiments, the detection of jets is very important. The multiparticle events we used for our studies are, of course, not the same as the QCD jets that originate from a fragmenting quark or gluon. Yet, for the purpose of calorimetry they are very useful, since they represent a collection of particles that enter the calorimeter simultaneously. The composition of this collection is unknown, but the total energy is known. In the absence of a jet test beam, this is 
a reasonable alternative.

In Section 2, the instruments and the experimental setup in which the measurements were carried out are described, as well as the calibration and data analysis methods that were used. Experimental results are presented in Section 3.
In Section 4, we investigate and compare different methods  to measure the energy resolution of this calorimeter.
Conclusions from these studies are presented in Section 5.

\section{Equipment and measurements}
\vskip-5mm
\subsection{Detectors and beam line}
\vskip -5mm

\begin{figure}[htbp]
\epsfysize=7.9cm
\centerline{\epsffile{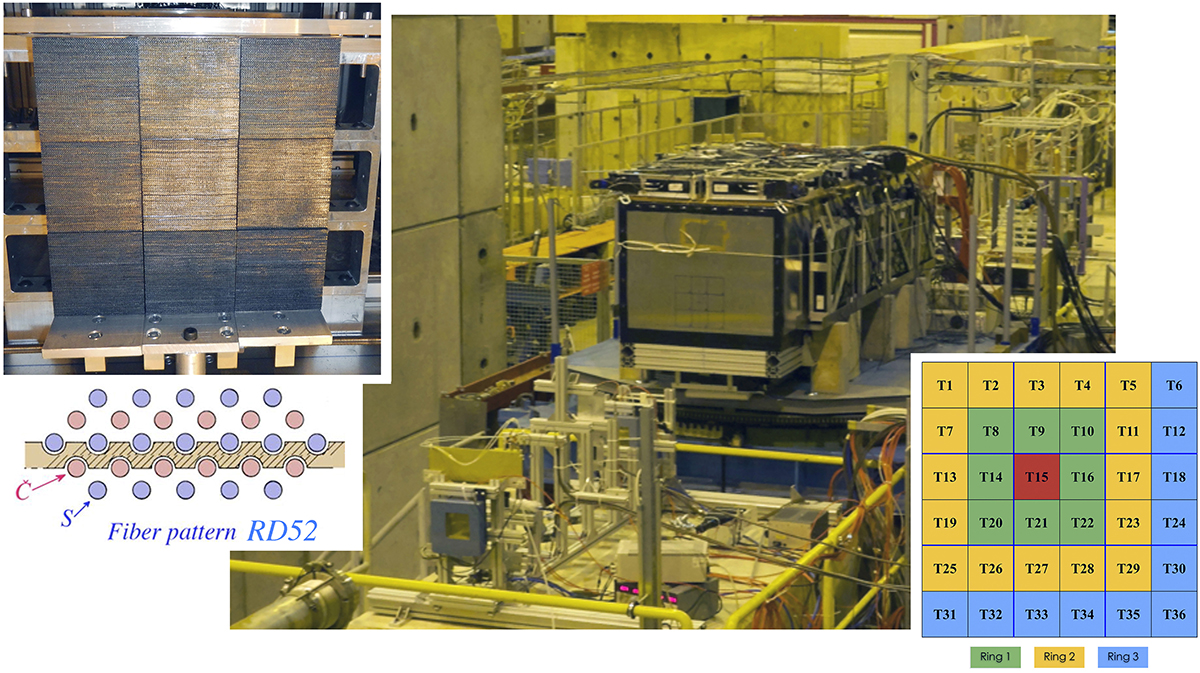}}
\caption{\footnotesize Experimental setup in the H8 beam of the SPS at CERN. The calorimeter is installed inside a light tight box, surrounded on 4 sides by 20 modular leakage counters. The entire setup is installed on a movable table, which allows the impact point and the incident angle of the beam particles to be chosen as needed.
The beam particles arrive through the vacuum pipe visible in the bottom left corner, and pass through several beam defining elements upstream of the calorimeter. The left insert shows a picture of the front face of the calorimeter, which consists of a 3$\times$3 matrix of modules, and the arrangement of the scintillating and \v{C}erenkov fibers in the lead absorber. 
The right insert shows the tower structure, one central tower surrounded by two complete rings and one incomplete one.}
\label{setup}
\end{figure}

For these particular studies, which were carried out in October 2015, we used secondary or tertiary beams derived from the 400 GeV proton beam delivered by the CERN Super Proton Synchrotron. These particle beams were steered through the H8 line into the RD52 fiber calorimeter. 
Figure \ref{setup} shows the experimental setup. 

The fiber calorimeter used for the studies described in this paper is modular, and uses lead as the absorber material.
Each of the nine modules is 2.5 m long (10 $\lambda_{\rm int}$), has a cross section of $9.2 \times 9.2$ cm$^2$ and a fiducial mass of 150 kg. Each module consists of four towers
(4.6$\times$4.6$\times$250 cm$^3$), and each tower contains 1024 plastic optical fibers (diameter 1.0 mm, equal numbers of scintillating and clear plastic fibers)\footnote{The scintillating fibers are of the SCSF-78 type, produced by Kuraray, the \v{C}erenkov light is generated in PMMA-based SK40 fibers, produced by Mitsubishi.}. Each tower produces two signals, a scintillation signal and a \v{C}erenkov signal, which are detected by separate PMTs\footnote{10-stage Hamamatsu R8900 and R8900-100. In order to limit the effects of self absorption on the signals, the PMTs detecting scintillation light are equipped with yellow filters.}.
For this reason, this type of detector is also known as a DREAM (Dual-REAdout Method) calorimeter. The sampling fraction for minimum ionizing particles in this calorimeter, both for the scintillation and for the \v{C}erenkov sampling structure, is 5.3\%. 

\begin{figure}[b!]
\epsfysize=5.2cm
\centerline{\epsffile{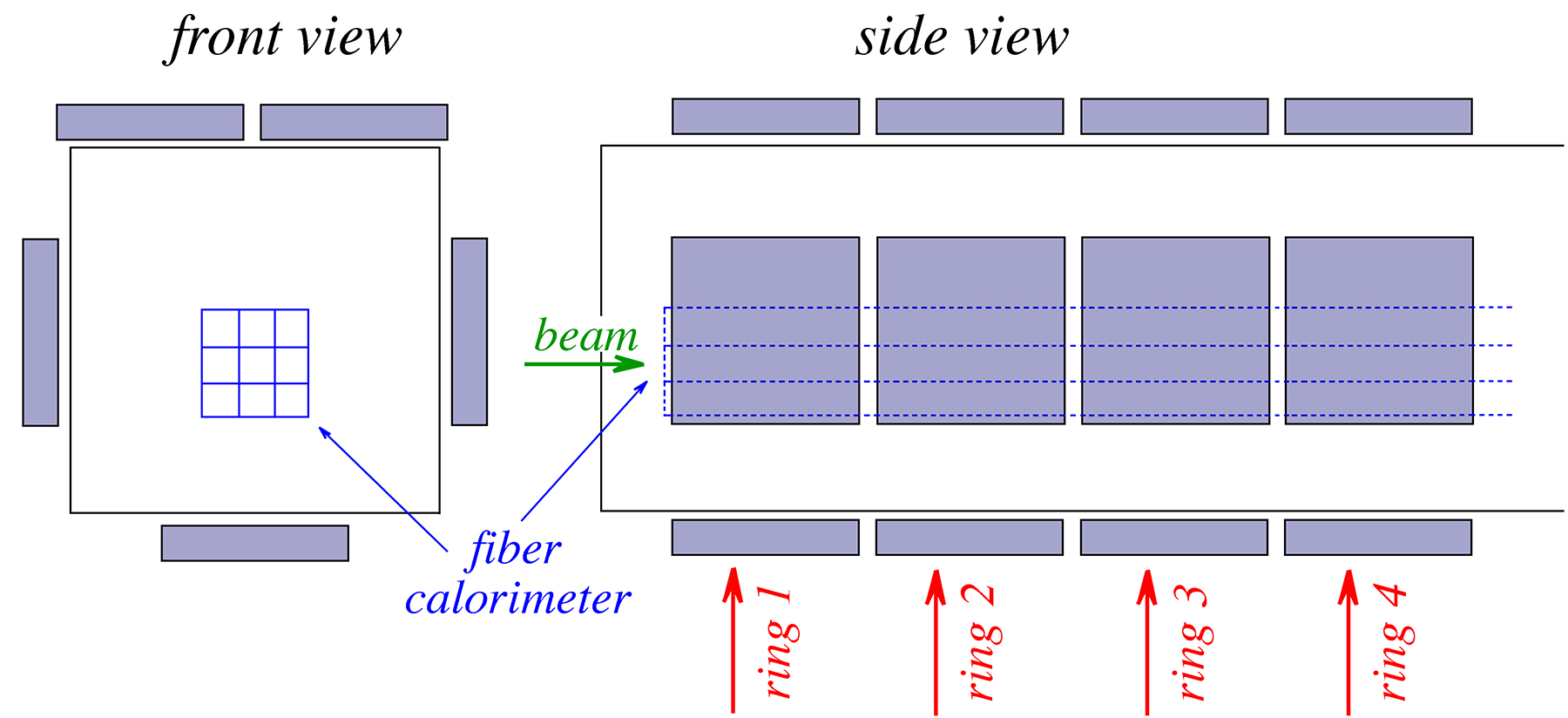}}
\caption{\footnotesize Location of the leakage counters with respect to the fiber calorimeter. The front view shows that five counters form
a ``ring" around the calorimeter, the side view shows that there are four such ``rings,''  located at different depths.}
\label{leakcounters}
\end{figure}

Measurements of the radial shower profile showed that the showers initiated by 60 GeV $\pi^-$ were, on average, contained at the level of $\sim 93\%$ in this structure. Electromagnetic showers are contained to better than 99\% and shower leakage was thus not an issue for electrons and photons.
In order to detect the hadronic shower leakage, the calorimeter was surrounded by large slabs of 
plastic scintillator (50$\times$50$\times$10 cm$^3$, mass 25 kg). Twenty such counters
were used in these tests. They can be seen in Figure \ref{setup} on the top, the bottom and the right hand side of the box containing the calorimeter. The location of the leakage counters with respect to the fiber calorimeter is shown in Figure \ref{leakcounters}.

The experimental setup also contained a number of auxiliary detectors, which were intended to limit and define the effective size of the beam spot, to determine the identity of individual beam particles, and to measure their trajectory. 
Figure \ref{layout} shows a schematic overview of the beam line, in which the positions of these auxiliary counters are indicated (not to scale): 
\begin{figure}[htbp]
\epsfysize=3.3cm
\centerline{\epsffile{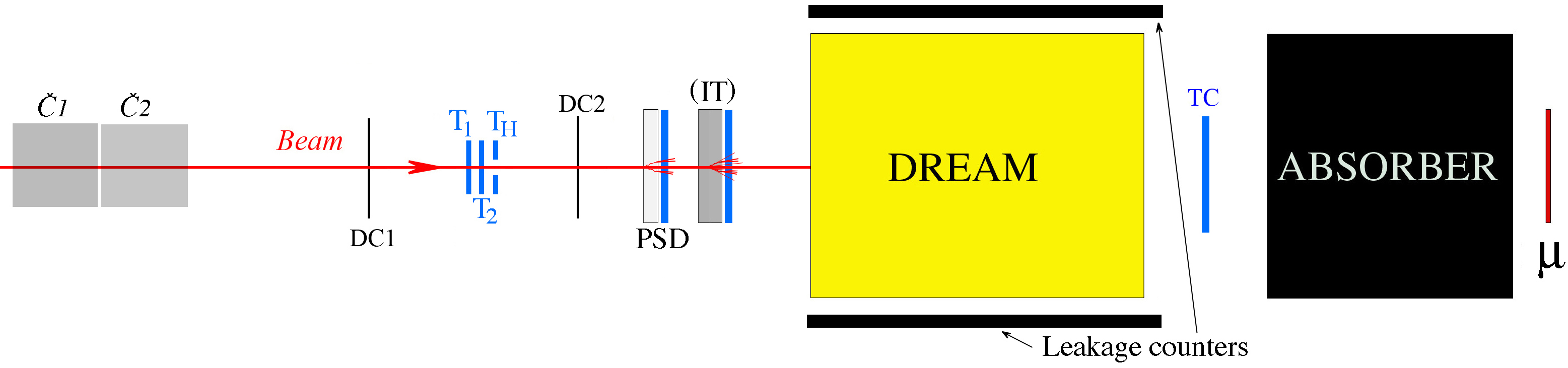}}
\caption{\footnotesize Schematic layout of the experimental setup (not to scale). Shown are the Threshold \v{C}erenkov counters (\v{C}1, \v{C}2), the delay wire chambers (DC1, DC2), the trigger counters (T$_1$, T$_2$, T$_H$), the preshower detector (PSD), the Interaction Target (IT), the calorimeter surrounded by leakage counters, the Tail Catcher (TC) and the Muon counter ($\mu$). See text for more details. }
\label{layout}
\end{figure}
\begin{itemize}
\item A set of three small scintillation counters provided the signals that were used to trigger the data acquisition system.
These trigger counters were 2.5 mm thick, the area of overlap between the first two ($T_1,T_2$) was 4$\times$4 cm$^2$. Downstream from these counters, a third scintillation counter ($T_H$) was installed. The latter had a hole with a radius of 10 mm in it. A (anti-)coincidence between the logic signals from these counters provided the trigger (${T_1} \cdot {T_2} \cdot {\overline{T_H}}$).

\item The trajectories of individual beam particles could be reconstructed with the information provided by two small delay wire chambers (DC1, DC2). This system made it possible to determine the location of the impact point of 80 GeV beam particles at the calorimeter surface with a precision of about 1 mm.

\item About 80 cm upstream of the calorimeter, a preshower detector (PSD) provided signals that could be used to remove electrons contaminating the hadron beams.
This PSD consisted of a 5 mm thick lead plate, followed by a 5 mm thick plastic scintillator. Electrons started developing showers in this device,
while muons and hadrons typically produced a signal characteristic of a minimum ionizing particle (mip) in the scintillator plate.

\item For certain (high) energies, an interaction target (IT), consisting of 10 cm of plastic, followed by a 5 mm thick plastic scintillator, was installed behind the PSD. This detector was used to create and select interactions in the plastic in which a significant number of secondaries were produced.
The signal in the scintillator provided a means to select events with a certain (minimum) multiplicity.
 
\item Downstream of the calorimeter, a Tail Catcher (TC) could also serve to help identify pions and muons.
This Tail Catcher consisted of a simple $20\times 20$ cm$^2$ scintillation counter. Electrons were fully absorbed in the calorimeter and thus did not create a signal in this detector, while muons produced a mip signal in it. Larger signals were typically caused by late showering hadrons.

\item Further downstream of the calorimeter, behind an additional $8 \lambda_{\rm int}$ worth of absorber, 
a 50$\times$50 cm$^2$ scintillation counter ($\mu$) served to identify muons that contaminated the particle
beam. 

\item About 50 m upstream of the calorimeter, two Threshold \v{C}erenkov counters (\v{C}$_{1,2}$) provided signals that made it possible to identify the type of beam particle. These counters were filled with CO$_2$ gas at a pressure that was chosen depending on the beam energy. These counters were in practice used to separate pions from protons.
\end{itemize}

The calorimeter was mounted on a table that could be displaced both horizontally and vertically, and also rotated around the vertical axis.
This allowed us to choose both the impact point and the angle of incidence of the beam particles. 

\subsection{Data acquisition}
\vskip -5mm
In order to minimize delays in the DAQ system, short, fast cables  were used to transport the signals from the trigger counters to the counting room. All other signals were transported through standard RG-58 cables with (for timing purposes) appropriate lengths.

In the counting room, signals from the Threshold \v{C}erenkov counters, PSD, Interaction Target, Tail Catcher and muon counter were fed into charge ADCs.
The signals from the wire chambers were fed into TDCs. 
The data acquisition system used VME electronics.
Two VME crates hosted all the needed readout and control boards.
The signals from the auxiliary detectors (Threshold \v{C}erenkov counters, PSD, Interaction Target, Tail Catcher, and Muon counter) were integrated and digitized with a sensitivity of 100 fC/count and a 12-bit dynamic range  on a 32-channel CAEN V862AC module. The timing information of the tracking chambers was recorded with 1 ns resolution in a 16-channel  CAEN V775N TDC, and was converted into ($x,y$) coordinates of the point where the beam particle traversed the chamber. 

Our readout scheme optimized the CPU utilization and the data taking efficiency using the bunch structure of the SPS accelerator cycle (which lasted between 36 and 54 s, depending on the various tasks of the accelerator complex), during which period beam particles were provided to our experiment by means of two extractions with a duration of 4.8 seconds each.

\subsection{Experimental data, calibration and analysis methods}
\vskip-5mm

The measurements described in this paper were performed in the H8 beam line of the Super Proton Synchrotron at CERN. 
We used either secondary beams directly produced by the 400 GeV protons from the accelerator on a target shared by several
beam lines, or tertiary beams derived from these secondary ones. For the experiments described in this paper, a secondary beam
of 20 GeV positive particles was used to calibrate all calorimeter towers. This beam consisted almost exclusively of positrons.
Secondary beams of 60 GeV and 180 GeV were used to measure the response functions at a variety of energies. Low energy
beams were derived as tertiaries from the 60 GeV secondary beam, and beams with energies above 60 GeV were derived from the 
180 GeV secondary beam. The latter was also used to provide 180 $\mu^+$ particles (obtained by blocking all other particles with absorbers), which were used to calibrate the leakage counters.

For the calibration runs, beam particles were steered into the center of each of the 36 individual calorimeter towers (see insert Figure \ref{setup}),  or through the central plane of the leakage counters. 
For each run, 10 000 events were collected, while 10\% randomly triggered events provided pedestal information.
The information from the wire chambers was used to select events in which the particles hit the calorimeter within a beam spot with a diameter of 10 mm. The HV settings were chosen such that the average calorimeter signal corresponded to several hundred ADC counts. 
The calibration runs were used to determine the energy equivalent of one ADC count for all individual signals, \ie the 36 scintillation signals, the 36 \v{C}erenkov signals and the 20 signals from the leakage counters. These calibration constants formed the basis for the energy determination of the hadronic events. The 72 calibration constants of the calorimeter towers were determined with 20 GeV positrons, which deposited 93\% of their energy in one individual tower.
The calibration constants of the leakage counters were equalized using 180 GeV muons which traversed the 50 cm long central plane of each counter. The overall scale of the leakage signals was set with 60 GeV pions sent into the 
center of the calorimeter. The total signal from all leakage counters combined was set to 3.84 GeV, representing 6.4\% of the particle energy.
In the analysis of the hadronic calorimeter performance described in the following sections, the scintillation signal is the sum of the signals measured in the scintillating fibers embedded in the calorimeter and the signals from the leakage counters, both expressed in GeV, using the scale derived from the electron calibration. The electron scale was also used for the hadron signals from the \v{C}erenkov fibers.

Dedicated hadron runs were carried out for the following energies and polarities: +20, +40, $\pm 60$, +80, +100, and +125 GeV. Dedicated multiparticle  (``jet'') runs were performed at +40, +60, +100 and +125 GeV.
For each event selected by the trigger counters, the ADC  data from the auxiliary detectors and the TDC data from the wire chambers were recorded.
Off-line, the beam chamber information was used to select events within a small beam spot (typically with a radius $< 5$~mm). The information provided by the auxiliary detectors was used to identify and select the desired particles.

In order to select hadron event samples, the electrons and muons had to be removed from the collected events. This had to be done in a way
that would not bias the resulting hadron event samples, and therefore had to be based entirely on the auxiliary detectors.
Electrons (or positrons) were identified as particles that produced a signal in the PSD that was larger than $\sim 200$ ADC counts above pedestal, which corresponds to the combined signals produced by two minimum ionizing particles (mips) traversing this detector. Additional requirements were that no signals incompatible with electronic noise were produced in the muon counter. 
Muons were identified as particles that produced a signal incompatible with electronic noise in the muon counter. At low energies, a significant fraction of the muons did not traverse that counter because of multiple scattering (or absorption) in the upstream material. In that case, particles were also identified as muons if they produced signals in the PSD or IT, as well as in the Tail Catcher, that were compatible with a mip, and no signal incompatible with the pedestal in the sum of all leakage counters.\hb
Protons were defined as particles that produced a signal compatible with the pedestal in both upstream Threshold \v{C}erenkov counters.
Pions were required to produce a signal in at least one of these counters that was significantly (at least 3$\sigma$) above pedestal.
Table \ref{ppi} lists the percentages of protons and pions in the hadron event samples determined on the basis of this criterion, as well as the percentage of contaminating electrons and muons.

\begin{figure}[htbp]
\epsfysize=7.5cm
\centerline{\epsffile{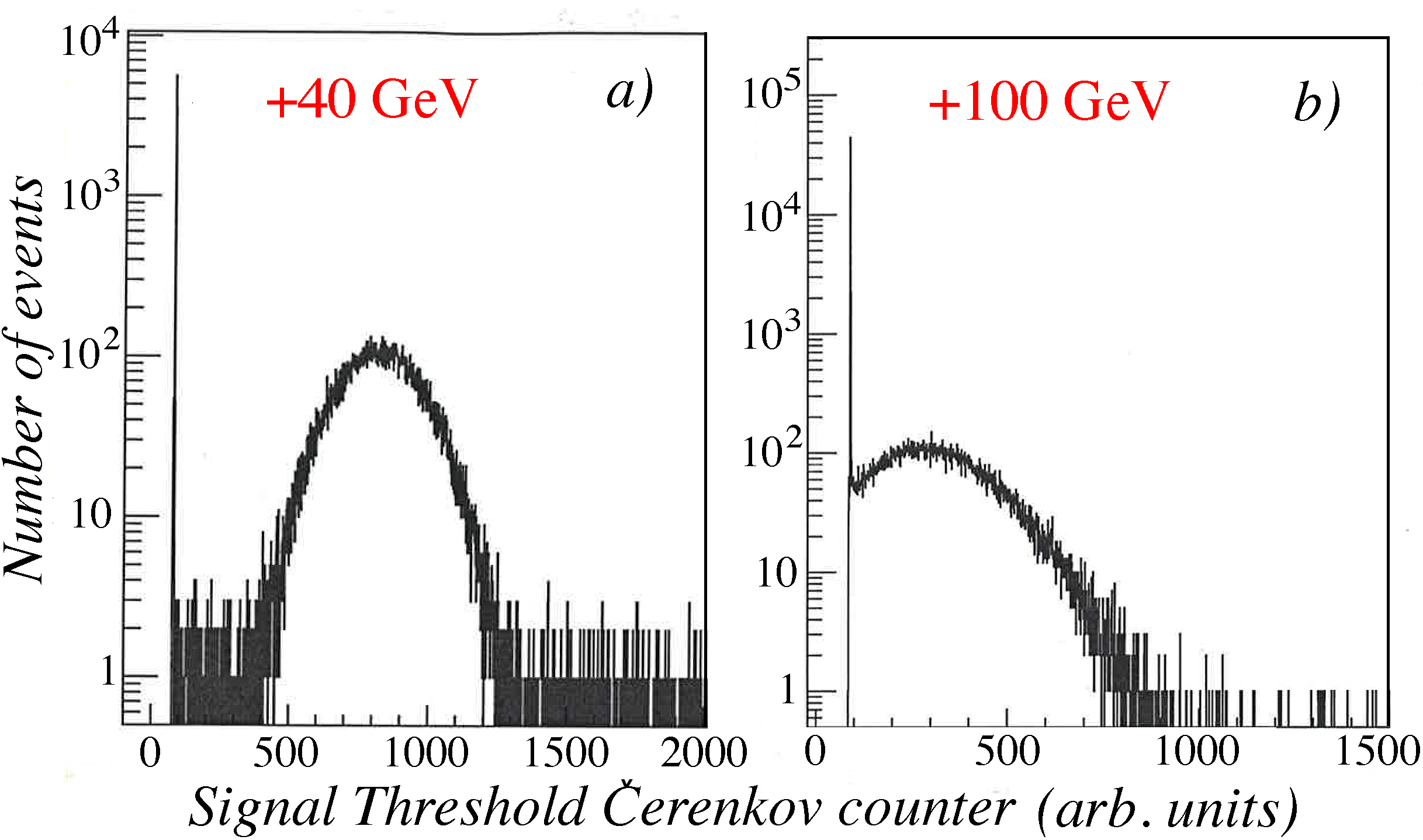}}
\caption{\footnotesize Signal distributions in one of the Threshold \v{C}erenkov counters for positive particles of 40 GeV ($a$) and 100 GeV ($b$). The gas pressure was such that protons would {\sl not} produce a signal, but pions would.}
\label{Ccounter}
\end{figure}
\begin{table}[hbtp]
\caption{\footnotesize Percentage of events that qualified as electrons, muons, pions and protons. The proton percentage is a maximum. It contains a possible contamination (misidentified pions) that is listed as well.}
\centering
\vskip 3mm
\begin{tabular} {|cccccc|} \hline
{\sl Beam energy}&{\sl Electrons}&{\sl Muons}&{\sl Pions}&{\sl Protons}&{\sl Possible contamination of $p$}  \\ \hline
+10 GeV&62\%& --&38\%   &--& --  \\
+20 GeV&31\%& 0.9\%&58\%   &10\%& --  \\
+40 GeV&41\%& 1.9\%&41\%   &16\%& 3\%  \\
%-40 GeV&?\%& --&?\%   &0& --  \\
+60 GeV&30\%& 1.8\%&44\%   &24\%& 3\%  \\
-60 GeV&23\%& 0.9\%&76\%   &--& --  \\
+80 GeV&20\%& 2.7\%&36\%   &41\%& 4\% \\
%-80 GeV&?\%& --&?\%   &0& --  \\
+100 GeV&14\%& 4\%&20\%   &62\%& --  \\
+125 GeV&5\%& 40\%&10\%   &45\%& --  \\
+180 GeV&--& 1\%&2\%   &97\%& --  \\ \hline
\end{tabular}
\label{ppi}
\end{table}

The counters were not fully efficient, and therefore the resulting proton/pion separation was not perfect, especially at the highest energies. Figure \ref{Ccounter} shows the signal distributions measured for beams of +40 GeV and +100 GeV. 

In order to determine the inefficiency of the Threshold \v{C}erenkov counters, we compared the signal distributions measured for beams with negative and positive polarity at the same energy. This comparison assumes that the production of antiprotons on the production target is negligible. Table \ref{ppbar} shows the fraction of hadrons (\ie after electrons and muons have been removed from the event samples) that produced signals compatible with the pedestal in {\sl both} Threshold \v{C}erenkov counters, for beams of 40, 60 and 80 GeV. 
\begin{table}[b!]
\caption{\footnotesize Percentage of hadronic events with pedestal signals  in {\sl both} Threshold
\v{C}erenkov counters.}
\centering
\vskip 3mm
\begin{tabular} {|cc|cc|} \hline
{\sl Beam energy}&{\sl pedestals in \v{C}$_1$,\v{C}$_2$}&{\sl Beam energy}&{\sl pedestals in \v{C}$_1$,\v{C}$_2$}  \\ \hline
+40 GeV&27.3\%   &-40 GeV& 5.1\%  \\
+60 GeV  &35.0\%   &-60 GeV& 3.6\%  \\ 
+80 GeV&53.5\%&-80 GeV& 4.6\%  \\ \hline
\end{tabular}
\label{ppbar}
\end{table}

Based on these considerations, we estimated the purity of the proton sample. The possible contamination of the proton sample by misidentified pions is listed in the last column of Table \ref{ppi}, as a percentage of the total number of events. The negative polarity hadrons were all considered pions. No attempts were made to measure the contribution of kaons to the various event samples. Such contributions are estimated (from beam simulations) to be at the few percent level. Since kaons would also generate
pedestal events in the Threshold \v{C}erenkov counters, the percentage of protons listed in Table 1 is a maximum.

No distinction was made between protons and pions for the measurements with the Interaction Target. Interacting hadrons were selected by means of a cut in the signal from the scintillation plate connected to the downstream end of this plastic target. An interacting hadron was defined as an event in which a signal compatible with a mip was produced in the PSD, combined with a signal larger than a certain minimum value (equivalent to 6 mips) in the IT.

\section{Experimental results}
\vskip -5mm
\subsection{The Dual Readout Method}
\vskip-5mm
The Dual-Readout approach for measuring hadron showers exploits the fact that the energy carried by the non-em shower component  of hadron showers is mostly deposited by non-relativistic shower particles (protons), and therefore does not contribute to the signals of a \v{C}erenkov calorimeter. By measuring simultaneously the visible deposited energy ($dE/dx$) and the \v{C}erenkov light generated in the shower absorption process, one can determine $f_{\rm em}$ {\em event by event} and thus eliminate
(the effects of) its fluctuations. The correct hadron energy can be determined from a combination of both signals.

This principle was first experimentally demonstrated by the DREAM Collaboration \cite{Akc05a}, with a Cu/fiber calorimeter. Scintillating fibers measured $dE/dx$, and quartz fibers measured the \v{C}erenkov light.
The response ratio of these two signals was related to $f_{\rm em}$ as
\begin{equation}
{C\over S} ~=~ {f_{\rm em} + 0.21~(1 - f_{\rm em})\over {f_{\rm em} + 0.77~ (1 - f_{\rm em})}}
\label{eq1}
\end{equation}
where 0.21 and 0.77 represent the $h/e$ ratios of the \v{C}erenkov and scintillator calorimeter structures, respectively.  The hadron energy could be derived directly from the two signals \cite{deg07}:
\begin{equation}
E~=~{{S - \chi C}\over{1 - \chi}}~,~~~~~{\rm with}~~~\chi = {{\Bigl[1- (h/e)_S\Bigr]}\over{\Bigl[1 - (h/e)_C\Bigr]}} ~\approx ~0.3
\label{eq2}
\end{equation}
The $e/h$ values, and thus the value of the parameter $\chi$ are a bit different when lead absorber is used. 

\subsection{Impact of the leakage counters}
\vskip -5mm
\begin{figure}[htbp]
\epsfysize=6cm
\centerline{\epsffile{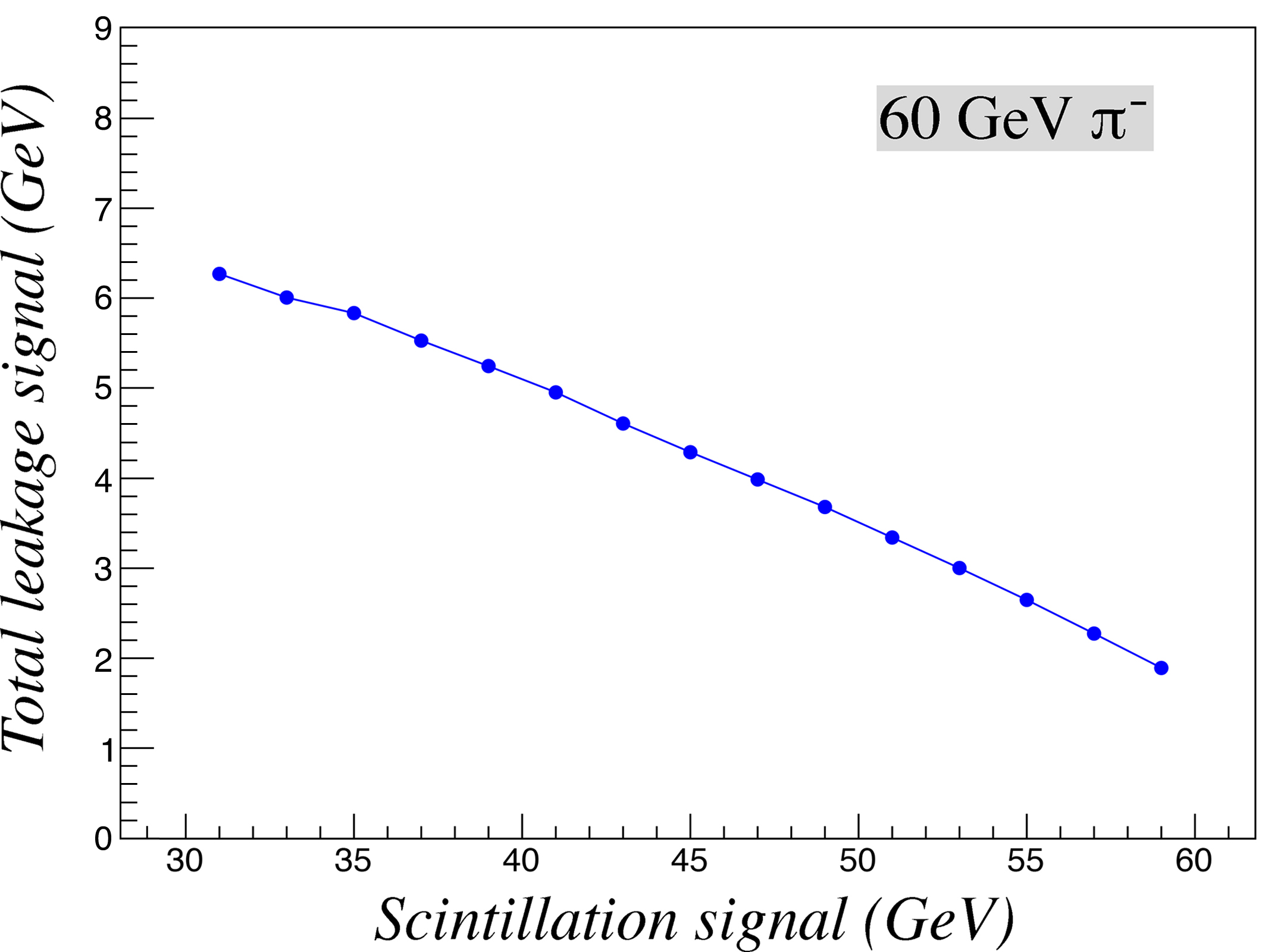}}
\caption{\footnotesize Relationship between the average signals from 60 GeV $\pi^-$ showers measured in the scintillation channels of the calorimeter and in the array of leakage counters. }
\label{LvsS}
\end{figure}
In order to study the effectiveness of the described leakage counters, we first studied the correlation between the signals from these counters and the scintillation signals from the fiber calorimeter. The result, shown in Figure \ref{LvsS} for 60 GeV $\pi^-$, indicates that there is indeed a good anti-correlation between the {\sl average} signals. However, the resolution improvement depends of course on the {\sl event-by-event anti-correlation}. The counters turned out to be indeed somewhat effective in that respect. 

\begin{figure}[b!]
\epsfysize=9cm
\centerline{\epsffile{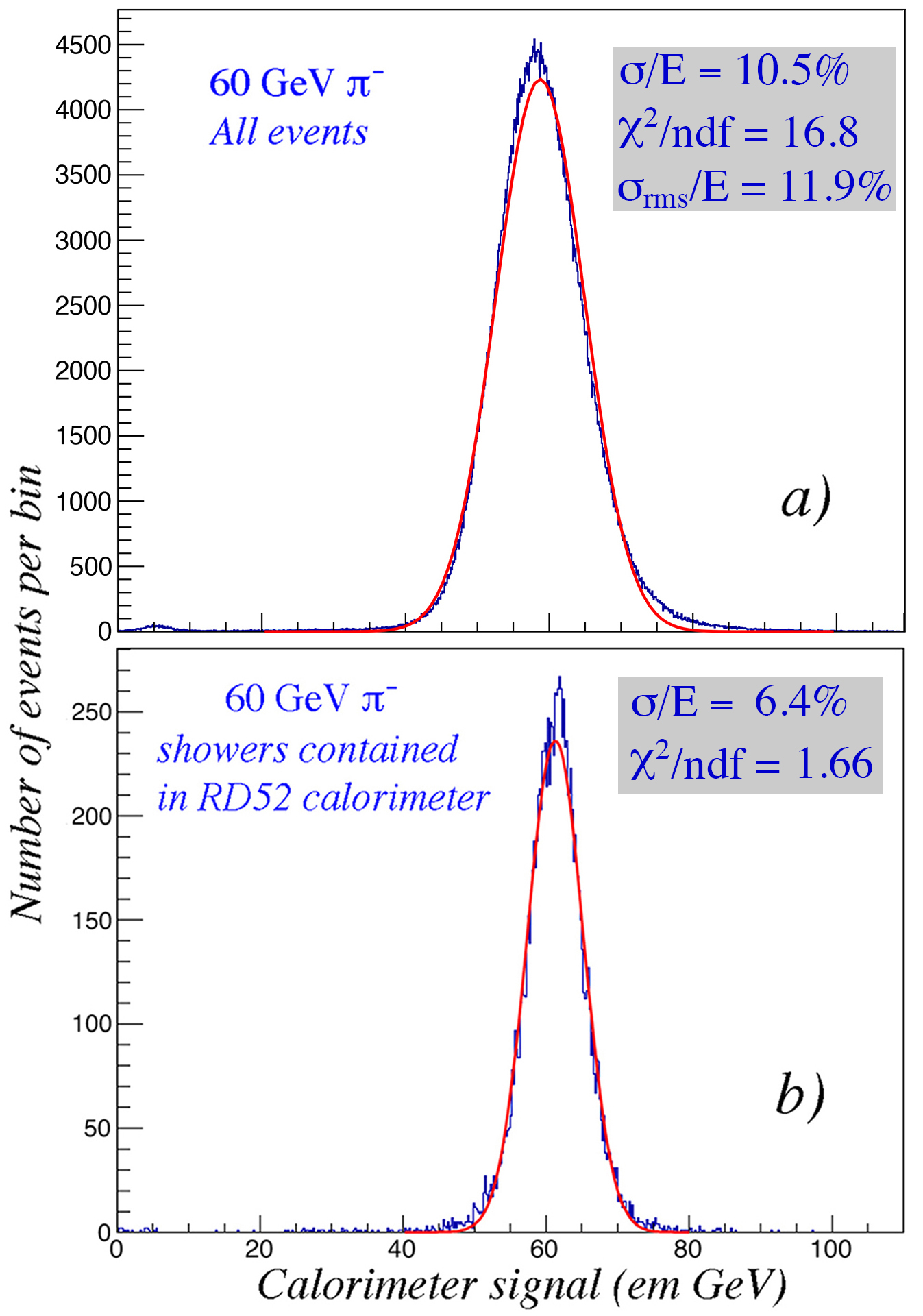}}
\caption{\footnotesize Total signal distributions for 60 GeV $\pi^-$, measured with the dual-readout method. Shown are the distributions for all events ($a$) and for events that were fully contained inside the calorimeter, \ie for which no energy leakage was measured in the leakage counters ($b$).}
\label{pi60cont}
\end{figure}

An extreme example of this effectiveness is shown in Figure \ref{pi60cont}, in which the signal distribution for all events (Figure \ref{pi60cont}a) is compared with the signal distribution 
for the events in which {\sl no} shower leakage was observed, \ie the (small fraction of the) events that were entirely contained in the fiber calorimeter. The latter distribution exhibits an energy resolution that is almost a factor of two better, and is in addition well described by a Gaussian function. These signal distributions were obtained with the standard dual-readout procedure (Section 3.1).

The signal distribution for all 60 GeV $\pi^-$ events shows deviations from a Gaussian shape. The type of deviations indicates that effects of light attenuation in the (scintillating) fibers are responsible for this \cite{Aco90}. The response of the fibers is not uniform in depth. Because of light attenuation, the response gradually increases as the light is produced closer to the PMTs, \ie deeper inside the calorimeter. The convolution of the attenuation curve with the longitudinal light production profile in hadron showers leads
to a response function with the measured characteristics.
The steeper the light attenuation curve, the more pronounced these effects become. It has been demonstrated that the effective light attenuation length increases with the distance to the light detector \cite{Hartjes}. An important feature responsible for this is the ``cladding light,'' which is much stronger attenuated than light trapped inside the fiber core. Therefore, this cladding light contributes predominantly to the signals from energy deposited close to the light detector. By making the upstream end of the fibers reflective, the attenuation curve becomes flatter,   
increasingly so as the distance to the light detector increases. In this way, effective attenuation lengths in excess of 8 m were obtained for the fibers in the SPACAL calorimeter \cite{Aco90}. However, in our calorimeter, the open end of the fibers was not made reflective, and the
attenuation length is therefore shorter.
The effect of this is an additional contribution to the hadronic energy resolution which, in first approximation, is energy independent.
This contribution, as well as the asymmetry of the response function, increases as the light is produced closer to the PMTs, \ie deeper inside the calorimeter.  
\begin{figure}[b!]
\epsfysize=9.5cm
\centerline{\epsffile{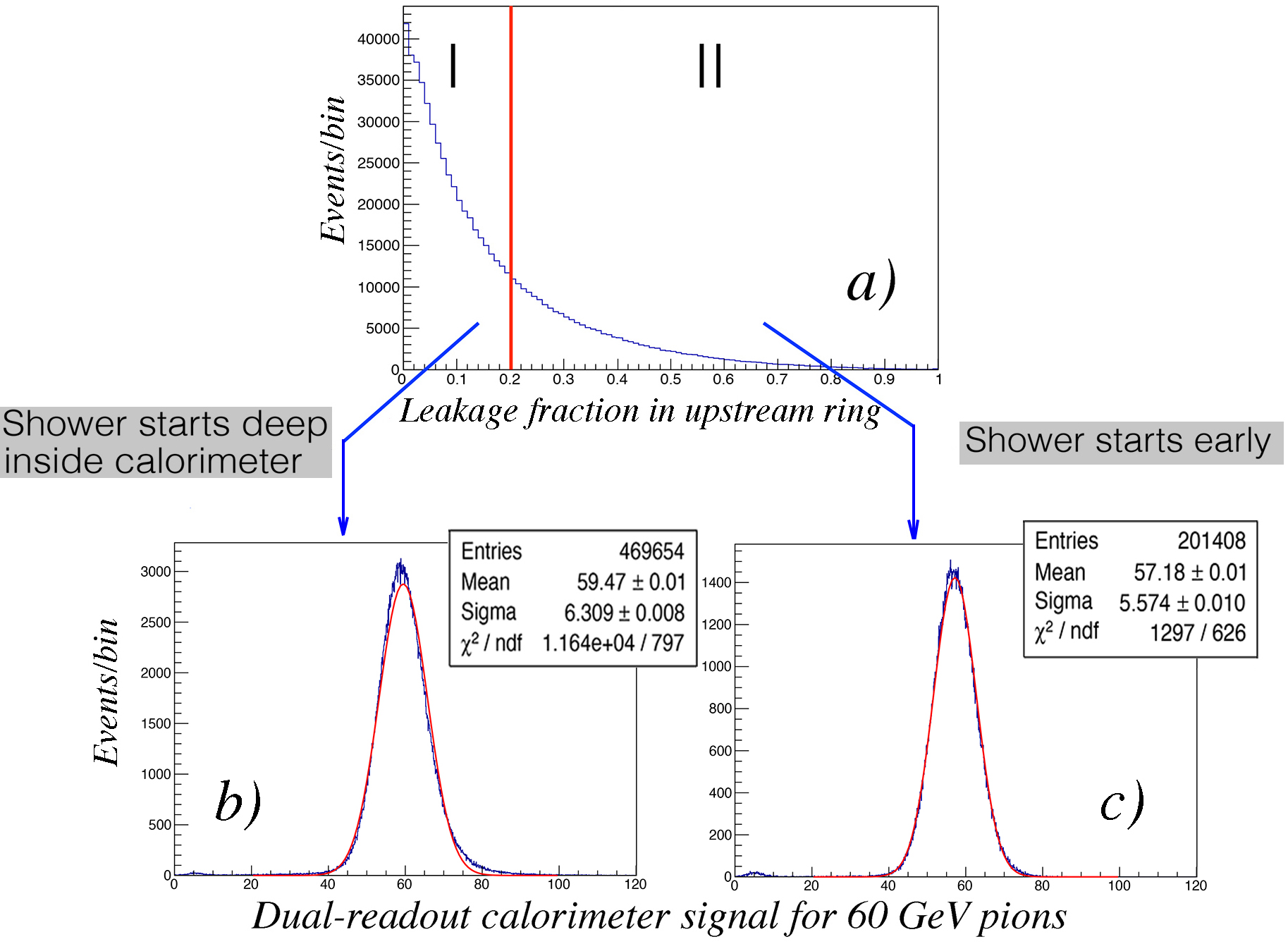}}
\caption{\footnotesize Signal distributions for 60 GeV $\pi^-$ measured with the dual-readout method ($b$,$c$), for different cuts on the fraction of the total leakage signal that was recorded in the first ring of  the leakage counter array ($a$). The energy scale is expressed in units of GeV, as established in the calibration with electrons. See text for more details.}
\label{Lcuts}
\end{figure}

To investigate these phenomena, we separated the events into sub-samples, based on the fraction of the total leakage signal that was measured in ring 1 of the leakage counters (see Figure \ref{leakcounters}). 
Figure \ref{Lcuts}a shows the distribution of that fraction. In general, we may assume that a small fraction indicates that the scintillation light is, on average, produced deep inside the calorimeter, \ie in the region
where the light attenuation curve is steeper than for light produced close to the calorimeter's front face \cite{Hartjes}. 

Figure \ref{Lcuts}b shows that the asymmetry is indeed predominantly observed for events in which that fraction is small ($< 0.2$), \ie events in which most of the energy was deposited deep inside the calorimeter, where the effects of light attenuation are largest. The signal distributions for the other events are much more Gaussian (Figure \ref{Lcuts}c). The average calorimeter signal is also somewhat smaller for these events, which is consistent with larger attenuation losses due to the longer path length of the light on its way to the PMT.
These results indicate that light attenuation in the scintillating fibers was indeed a significant factor contributing to the hadronic energy resolution of this calorimeter. By comparing Figures \ref{pi60cont}a and \ref{pi60cont}b, one may conclude that leakage fluctuations contributed the rest.
\begin{figure}[htbp]
\epsfysize=5.5cm
\centerline{\epsffile{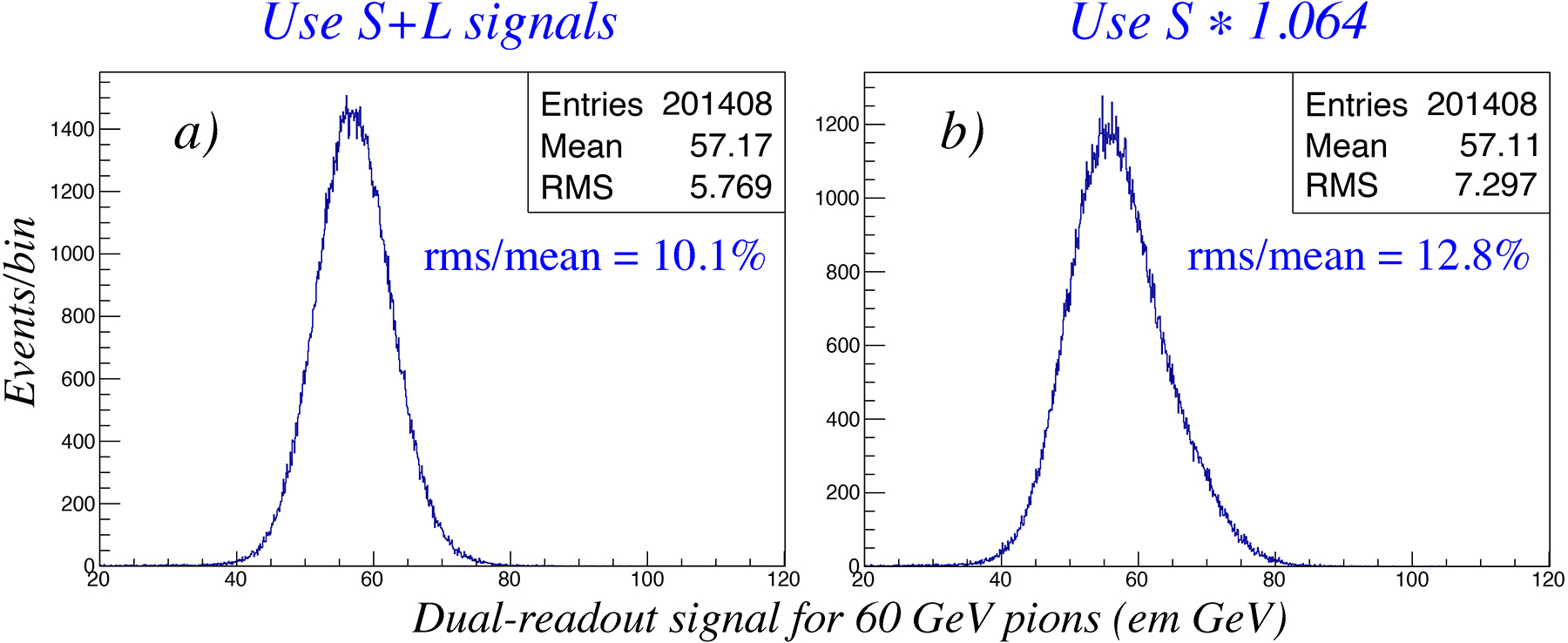}}
\caption{\footnotesize Signal distributions for 60 GeV $\pi^-$, measured with the dual-readout method. The scintillation signal from the fibers was increased with the total signal measured in the leakage counters ($a$), or simply multiplied by a factor of 1.064 for all events ($b$). }
\label{Leffect}
\end{figure}

In order to investigate how effective the signals from the leakage counters were in reducing the effects of side leakage on the energy resolution, we compared the signal distributions for the 60 GeV pions in which the leakage signals were added event by event to those from the scintillating fibers (Figure \ref{Leffect}a) with the signal distributions in which the fiber scintillation signals were all multiplied by a constant factor, representing the {\sl average} leakage fraction (Figure \ref{Leffect}b). The leakage counters did indeed improve the hadronic energy resolution significantly, albeit not as much as one might expect from a sufficiently enlarged fiber calorimeter. 
If we take the energy resolution of 6.4\% (Figure \ref{pi60cont}b) as the value for fully contained showers\footnote{GEANT4 based Monte Carlo simulations gave similar resolution values \cite{geant}.}, then the contribution of leakage fluctuations to the resolution shown in Figure \ref{Leffect} was reduced from 11.1\% in the absence of any leakage detection (Figure \ref{Leffect}b) to 7.8\% for the imperfect leakage detector used in these studies (Figure \ref{Leffect}a).

We have shown in the past that the light attenuation effects can be eliminated event by event through the time structure of the signals \cite{Akc14}, or by placing the calorimeter at a small ($\sim 1^\circ$) angle with the beam line \cite{Akc05a}. That information was not available during these tests. Instead, we have chosen to limit the analyses to event samples in which more than 20\% of the leakage signal was recorded in the first ring of the leakage counters. Figure \ref{Lcuts}c shows that this choice effectively eliminates events in which the hadrons start showering deep inside the calorimeter. 

\subsection{Proton/pion differences}
\vskip -5mm

A second issue we wanted to investigate with the data taken during the 2015 test beam period concerned the separation of pions and protons using {\sl only} calorimeter information. To that end, we used the two threshold \v{C}erenkov counters that were installed about 50 m upstream of the calorimeter setup. These counters were filled with CO$_2$ gas at a pressure that was chosen depending on the beam energy. 
Protons were defined as particles that produced a signal compatible with the pedestal in both \v{C}erenkov counters.
Pions were required to produce a signal significantly (at least $3\sigma$) above pedestal in at least one of these counters (see Figure \ref{Ccounter}).

In 1998, Akchurin and coworkers showed that there are significant differences between the shower development of high-energy protons and pions, which have measurable consequences for the signals from non-compensating 
calorimeters \cite{Akc98}. In prototype studies of the Forward Calorimeter for CMS, which is based on the detection of \v{C}erenkov light, they found that the signals from pions were typically $\sim 10\%$ larger than those from protons of the same energy. On the other hand, event-to-event fluctuations in these signals were $\sim 10\%$ smaller for protons, and the signal distributions were also more symmetric for protons.
These differences are a consequence of the requirement of baryon number conservation, which prohibits a $\pi^0$ from being the leading particle in proton induced showers.
\begin{figure}[htbp]
\epsfysize=10cm
\centerline{\epsffile{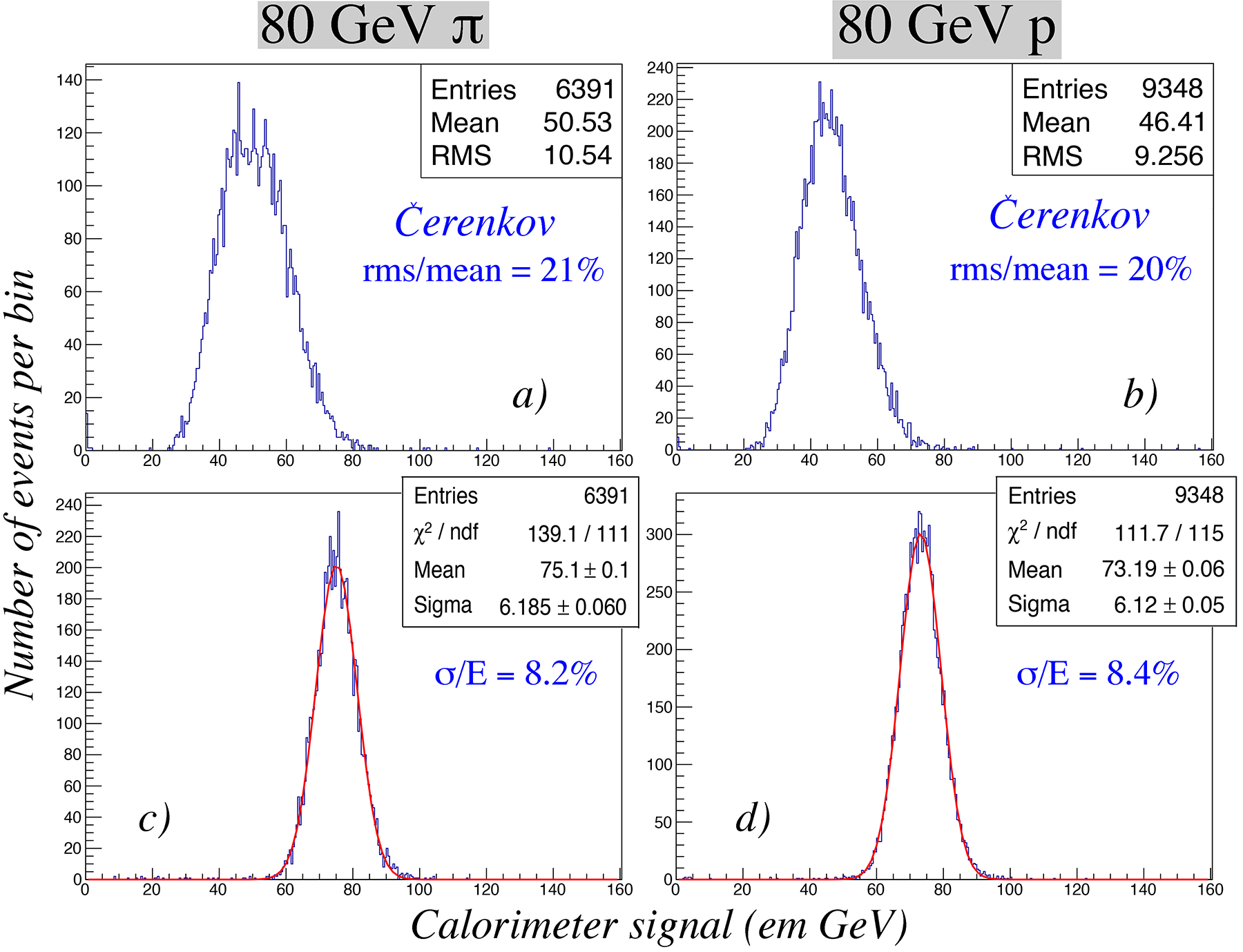}}
\caption{\footnotesize Signal distributions for 80 GeV pions and protons measured with the 9-module lead-based fiber calorimeter. Shown are the distributions for the \v{C}erenkov signals from 80 GeV $\pi^+$ ($a$) and protons ($b$), as well as the dual-readout total signals for 80 GeV $\pi^+$ ($c$) and protons ($d$). }
\label{ppicomp}
\end{figure}

Our data are in agreement with these findings. Figure \ref{ppicomp} shows signal distributions for the \v{C}erenkov signals from 80 GeV pions (\ref{ppicomp}a) and 80 GeV protons (\ref{ppicomp}b), respectively.
Indeed, the proton signals are, on average, $\sim 10\%$ smaller than the pion ones. On the other hand, the signal distribution for the protons is more symmetric and also somewhat narrower than the pion one.
Interestingly, 
a comparison between Figures \ref{ppicomp}c and \ref{ppicomp}d shows that application of the dual-readout method (Equations \ref{eq1},\ref{eq2}) largely eliminated the differences between these two types of showers. 
 
\subsection{The calorimeter performance for single hadrons}
\vskip -5mm

In this section, we present results on the energy resolution measured for single hadrons of different energies. For the positive polarity, separate samples of protons and pions were used.
No attempts were made to isolate the kaons, whose showers should also be different from pion ones in terms of the em shower component. Strangeness conservation prevents the production of leading $\pi^0$s in kaon induced showers, and therefore the characteristics of the em shower component (average value, event-to-event fluctuations in $f_{\rm em}$) are probably similar to those in proton induced showers. 
%In order to reduce the effects of light attenuation in the (scintillating) fibers, the samples were limited to events in which the signal in the most upstream ring of leakage counters represented more than 20\% of the total leakage signal. This eliminated events in which the hadrons started showering deep inside the calorimeter structure (see Figure \ref{Lcuts}).

For every event, two signals were available, a \v{C}erenkov signal and a scintillation signal. 
%The latter signal consisted of contributions from the scintillating fibers in the calorimeter and the leakage counters surrounding the calorimeter. 
The particle energy was found by combining these signals as in Equation \ref{eq2}, using a parameter value $\chi =0.45$. 
Both the reconstructed energy and the quality of the Gaussian fit are sensitive to the value of this parameter, and the chosen value represents 
the result of an optimization procedure in which $\chi$ was varied in small steps.
The optimal value is somewhat larger than for the copper-based DREAM calorimeter, as a result of differences between the $e/h$ values for lead  and copper, both for the scintillation and the \v{C}erenkov sampling structure \cite{Wig00}. 
We have used a value $\chi = 0.45$ for our lead-based calorimeter throughout this analysis. The procedures to obtain signal distributions for pions and protons were identical over the entire energy range studied here. 
\begin{figure}[htbp]
\epsfysize=4.6cm
\centerline{\epsffile{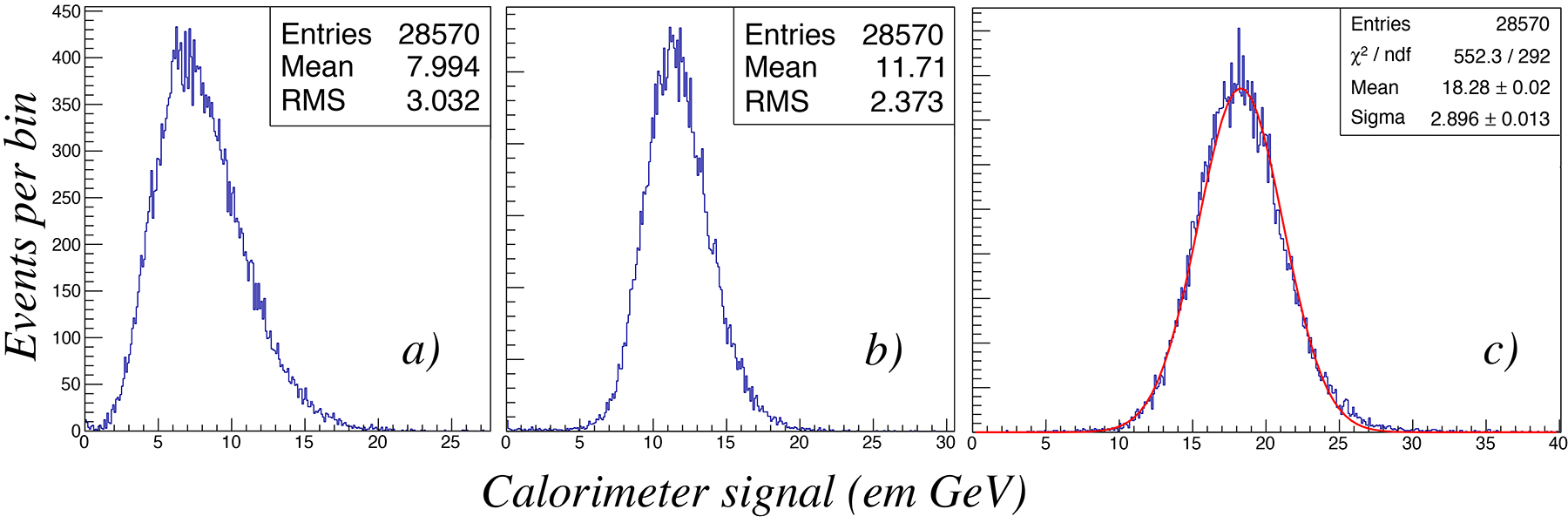}}
\caption{\footnotesize Signal distributions for 20 GeV $\pi^-$ particles. Shown are the measured \v{C}erenkov ($a$) and scintillation ($b$) signal distributions as well as the signal distribution obtained by combining the two signals according to Equation \ref{eq2}, using $\chi = 0.45$ ($c$). }
\label{pi-20}
\end{figure}

As an example, Figure \ref{pi-20} shows the distributions for the two individual signals, as well as the distribution of the dual-readout signals, combined according to Equation \ref{eq2}, for 20 GeV pions.
These distributions illustrate the benefits of the dual-readout method. Whereas the $C$ and $S$ distributions are rather wide and asymmetric, the dual-readout signal distribution is well described by a Gaussian fit.
\begin{figure}[b!]
\epsfysize=6.5cm
\centerline{\epsffile{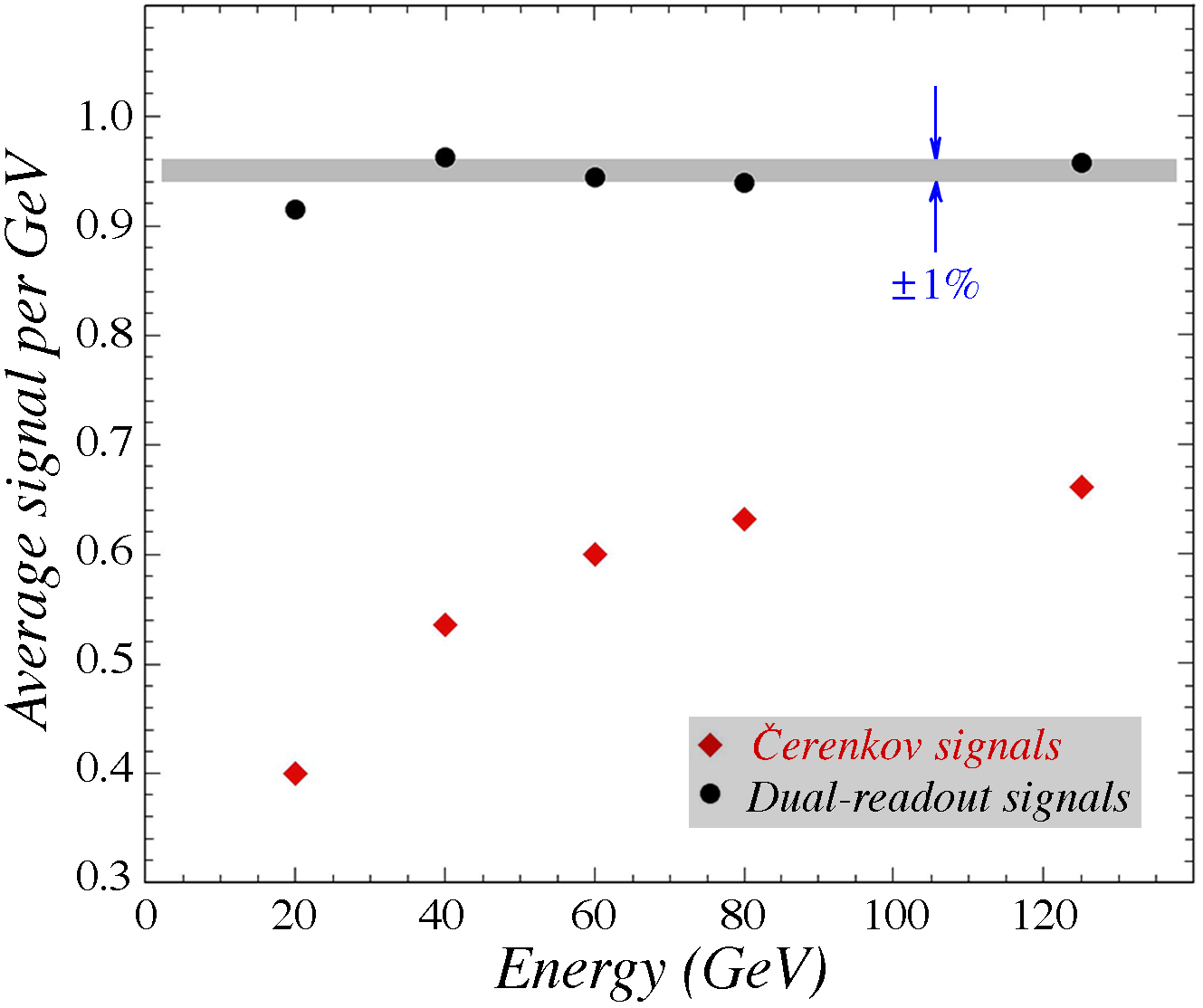}}
\caption{\footnotesize The hadronic response of the RD52 lead-fiber dual-readout calorimeter, for single pions. Shown are the average \v{C}erenkov signal and the dual-readout signal (Eq. \ref{eq2}) per unit deposited energy, as a function of the pion energy. }
\label{linhad}
\end{figure}
\begin{figure}[htbp]
\epsfysize=7.5cm
\centerline{\epsffile{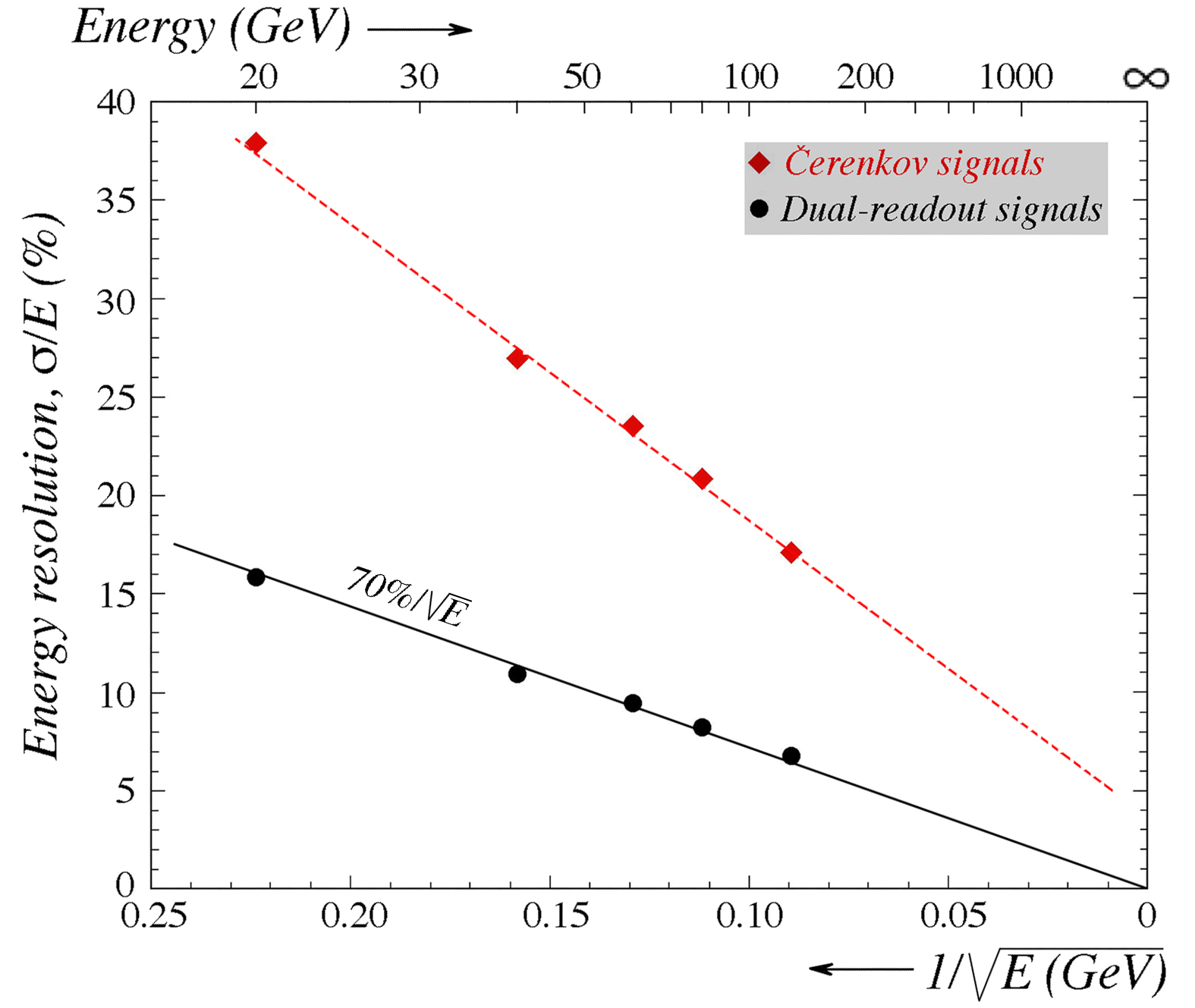}}
\caption{\footnotesize The hadronic energy resolution of the RD52 lead-fiber dual-readout calorimeter, for single pions. Shown are the results for the \v{C}erenkov signals alone, and for the dual-readout signals, obtained with Eq. \ref{eq2}.}
\label{reshad}
\end{figure}

The results of this study are summarized in Figures \ref{linhad} and \ref{reshad}. Figure \ref{linhad} shows the average signal per unit deposited energy
(\ie the {\sl calorimeter response}) as a function of energy, for pions with energies ranging from 20 to 125 GeV. Results are given separately for the \v{C}erenkov signals and for the dual-readout signals. Whereas the \v{C}erenkov response increased by more than 50\% over this energy range, the dual-readout response was constant to within a few percent, except for the lowest energy. The results for protons were essentially the same.

The hadronic energy resolution is shown as a function of energy in Figure \ref{reshad}. The energy scale is proportional to $-E^{-1/2}$, which means that the data points should be located on a straight line through the bottom right corner of this plot if the resolution is only determined by fluctuations
that are governed by Poisson statistics. Any deviation from such a line means that non-stochastic effects play a significant role. The experimental data show that the energy resolution for pions is well described by {\sl stochastic fluctuations alone} when the dual-readout signals are considered\footnote{Our Monte Carlo simulations have shown that this energy resolution is dominated by fluctuations in lateral shower leakage \cite{geant}.}.
On the other hand, the energy resolution measured on the basis of the \v{C}erenkov signals exhibits substantial deviations from $E^{-1/2}$ scaling. The straight line fit through the experimental data points suggests a 5\% resolution at infinite energy.
This is a consequence of the fact that the event-to-event fluctuations in the em shower fraction ($f_{\rm em}$) are not stochastic.

\section{The energy resolution of a calorimeter and how to determine it}
\vskip -2mm

\subsection{Introduction}
\vskip -5mm
In this section, we look in detail at the meaning of the term {\sl energy resolution} and discuss and compare various ways in which it is determined in practice. Strictly speaking, the energy resolution of a calorimeter describes the precision with which the energy of an unknown object that is absorbed in it can be determined. In practice, this important characteristic is usually measured with a beam of mono-energetic particles produced by an accelerator. This beam is sent into the detector. The (relative) energy resolution is deemed to be represented by the (fractional) width of the distribution of the signals produced by the calorimeter in response to these particles. Two important caveats should be mentioned in this context:
\begin{enumerate}
\item Since all particles typically enter the calorimeter in the same small area defined by the beam spot, the results are strictly speaking only valid for this particular part of the calorimeter. If the average signal varies with the impact point of the particles, which is often the case, then the real energy resolution is underestimated in this procedure.
\item The width of the signal distribution measured in this way is only indicative for the energy resolution if the average calorimeter signal indeed represents the correct energy of the beam particles. This condition may not be met, for example, when the calorimeter is intrinsically non-linear and has been calibrated at a different energy than that of the beam particles. It is also not met when the average signals are different for different types of particles with the same energy (\eg $\pi, K, p$) and the beam composition is unknown. 
\end{enumerate}
However, even when we assume that these caveats do not play a role for the calorimeter in question, one should realize that the procedure chosen to determine the energy resolution relies on the signal distribution from an ensemble of particles with the same energy.
This is a crucial aspect of methods that attempt to improve the energy resolution by techniques known as {\sl offline compensation}, or
{\sl Particle Flow Analysis}, where the width of a signal distribution is reduced with an iterative procedure, in which calibration constants of the various calorimeter sections that contribute to the signals are varied until an optimal result is obtained.

The procedure used to determine the energy resolution of the RD52 calorimeters, as described in the previous section, does {\sl not} rely on the availability of an ensemble of events created by particles of the same energy. The particle energy is determined with a simple formula (\ref{eq2}), which combines the values of two signals. The energy resolution is measured by comparing that calculated energy with the true energy of the particle that created the event. The availability of an ensemble of mono-energetic particles is not essential in this case.

However, if measuring the energy resolution is considered equivalent to measuring the width of the signal distribution for an ensemble of mono-energetic beam particles, then the dual-readout method also offers an alternative approach, described below. This approach leads to resolutions that are considerably better than the ones mentioned in the previous section. 

\subsection{The rotation method for single hadrons}
\vskip -3mm

\begin{figure}[htbp]
\epsfysize=10cm
\centerline{\epsffile{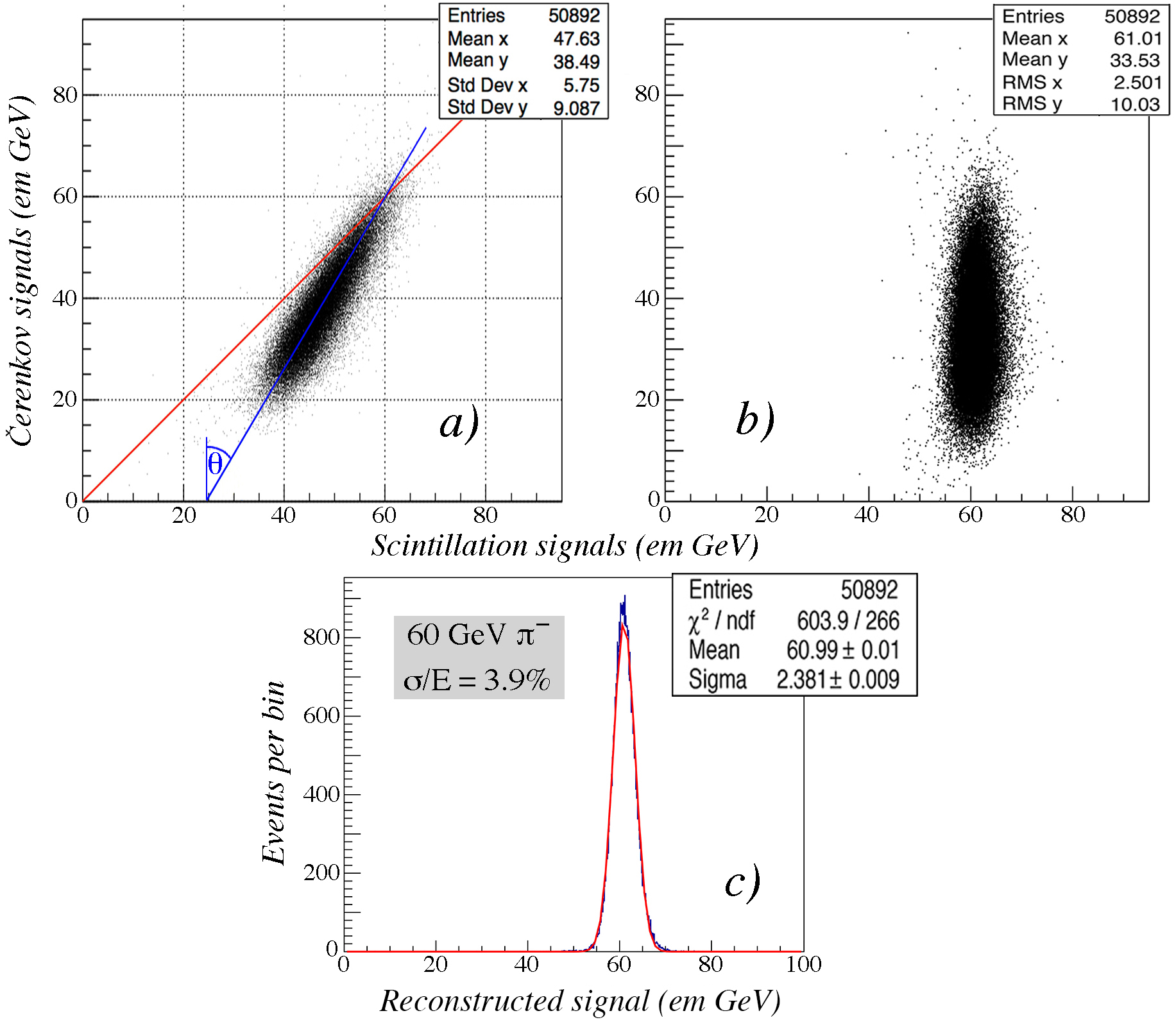}}
\caption{\footnotesize Signal distributions of the RD52 Dual-Readout lead/fiber calorimeter for 60 GeV pions. Scatter plot of the two types of signals as recorded for these particles ($a$) and rotated over an angle $\theta = 30^\circ$ around the point where the two lines from diagram $a$ intersect ($b$). Projection of the latter scatter plot on the $x$-axis ($c$).}
\label{DRrotation}
\end{figure}

Figure \ref{DRrotation}a shows a scatter plot of the \v{C}erenkov signals \vs the scintillation signals measured with this detector for 60 GeV pions. The signals from the leakage counters were added to those from the scintillating fibers, using the fact that the measured shower profile indicated that the side leakage at this energy was, on average, 6.4\%. The energy scale for both the \v{C}erenkov and the scintillation signals is given in units of GeV, derived from the calibration of these signals with electron showers.

This scatter plot shows the data points located on a locus, clustered around a line that intersects the $C/S = 1$ line at the beam energy
of 60 GeV. This is of course to be expected. In first approximation, the \v{C}erenkov fibers only produced signals generated by the electromagnetic components of the hadron showers, predominantly $\pi^0$s. The larger the em shower fraction, the larger the
$C/S$ signal ratio. Events in which (almost) the entire hadronic energy was deposited in the form of em shower components thus produced signals that were very similar to those from 60 GeV electrons and are, therefore, represented by data points located near (60,60) in this scatter plot.
The fact that the data points cluster around a straight line in this plot is in agreement with Groom's assessment of the fundamental aspects of dual-readout calorimetry \cite{PDG16}.

We can now rotate the scatter plot over the angle $\theta$  around this intersection point:
\begin{equation}
\left(\matrix{S^\prime \cr C^\prime\cr}\right)~=~
\left(\matrix{\cos{\theta}&\sin{\theta}\cr - \sin{\theta}&\cos{\theta}\cr}\right)
\left(\matrix{S\cr C\cr}\right)
\label{rot}
\end{equation}
and the result is shown in Figure \ref{DRrotation}b, for $\theta = 30^\circ$. The projection of this rotated scatter plot on the $x$-axis is shown in Figure \ref{DRrotation}c.
This signal distribution is well described by a Gaussian function with a central value of 61.0 GeV and a relative width, $\sigma/E$, of $3.9\%$. This corresponds to $30\%/\sqrt{E}$. The narrowness of this distribution reflects the clustering of the data points around the axis of the locus in Figure \ref{DRrotation}a.\index{Energy resolution!measured with test beams}
\begin{figure}[htbp]
\epsfysize=9cm
\centerline{\epsffile{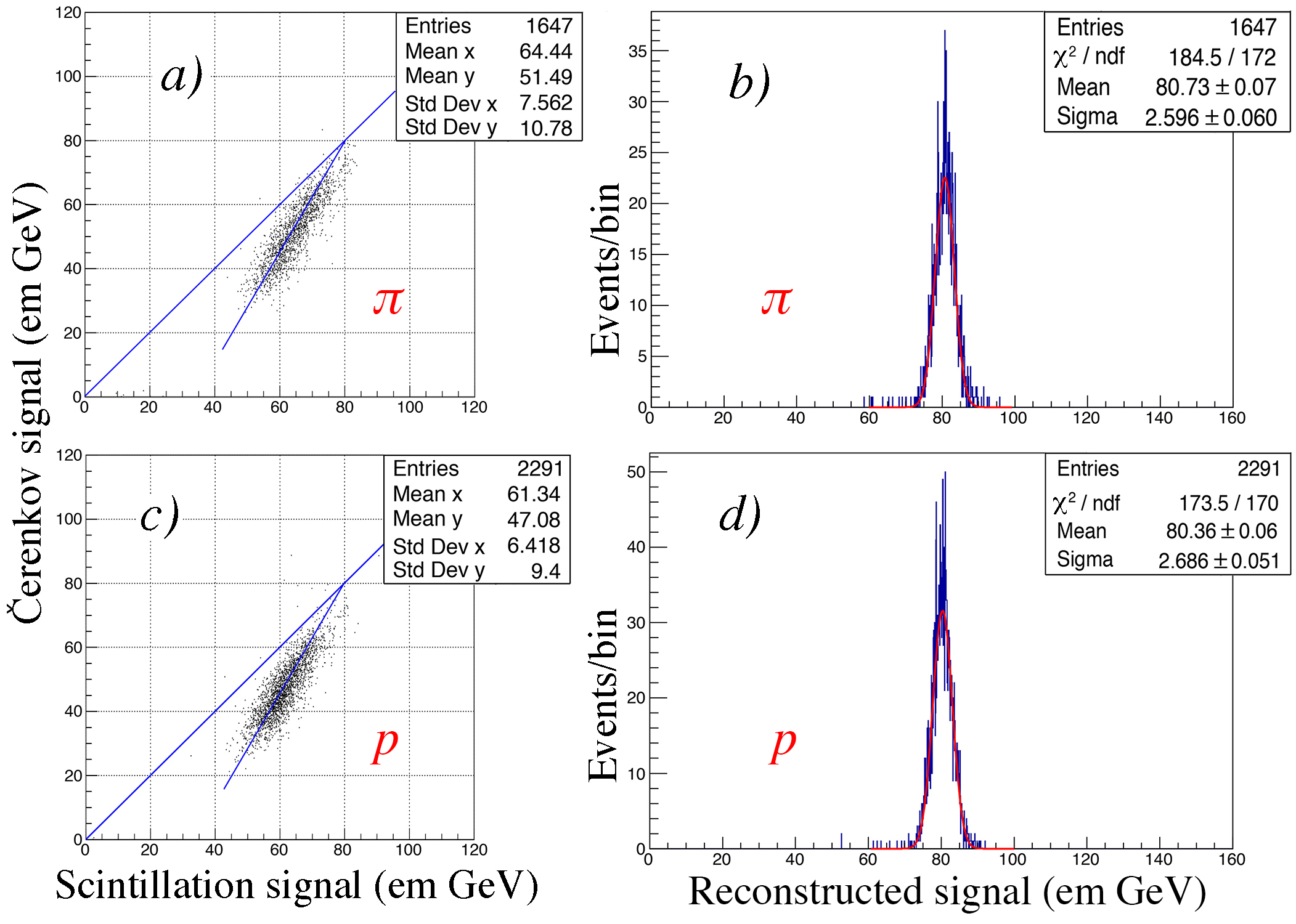}}
\caption{\footnotesize{Scatter plots of the \v{C}erenkov \vs the scintillation signals from showers induced by 80 GeV $\pi^+$ ($a$) and 80 GeV protons ($c$). Projection of the rotated scatter plots on the $x$ axis for the pions ($b$) and protons ($d$). The rotation procedure was identical to that used for 60 GeV $\pi^-$ (Figure \ref{DRrotation}).}}
\label{80ppi}
\end{figure}

We have applied exactly the same procedure for data taken at +125 GeV, +80 GeV, +40 GeV and +20 GeV, and obtained similar results. 
In addition, the use of positive polarity beams allowed us to separate the data into proton and $\pi^+$ samples.
Figure \ref{80ppi} shows the \v{C}erenkov \vs scintillation scatter plots for the 80 GeV $\pi^+$ (Figure \ref{80ppi}a) and proton (Figure \ref{80ppi}c) signals. These plots show a significant difference between the pion and proton signals. The average \v{C}erenkov signal is about 10\% larger for the pions than for the protons, a consequence of the absence of leading $\pi^0$s in the proton showers.
However, using the intersection of the axis of the locus and the $C/S = 1$ point as the center of rotation, and the same rotation angle ($30^\circ$) as for 60 GeV, the resulting signal distributions had about the same average value: 80.7 GeV for the pions (Figure \ref{80ppi}b) and 
80.4 GeV for the protons (Figure \ref{80ppi}d). The widths of both distributions were also about the same: 2.60 GeV for pions, 2.69 GeV for protons.
Regardless of the differences between the production of $\pi^0$s (and thus of \v{C}erenkov light) in these two types of showers,
the signal distributions obtained after the dual-readout procedure applied here, were thus practically indistinguishable.

We have applied exactly the same procedure for the 20 GeV, 40 GeV and the 125 GeV particles, with very similar results.
Also here, the average \v{C}erenkov signals in the raw data were significantly smaller for protons than for pions.
However, after applying the same rotation procedure as for the 60 and 80 GeV data (always using the same rotation angle, $\theta = 30^\circ$), the resulting signal distributions were centered around approximately the same values, and also the relative widths of these distributions were approximately the same.
The fact that the rotation angle used to achieve these results is independent of the particle type and the energy is consistent with Groom's observation that this angle only depends on the energy independent value of the $\chi$ parameter defined in Equation \ref{eq2} \cite{PDG16}.
\begin{table}[hbtp]
\caption{\footnotesize The reconstructed energy and the energy resolution for proton and pion showers, measured with the rotation method. See text for details.}
\centering
\vskip 3mm
\begin{tabular} {|lcccc|} \hline
{\sl Particles}&$\langle$\v{C} signal$\rangle$&$\langle$Reconstructed energy$\rangle$&$\sigma/E$ (\%)&${\sigma/E}\cdot \sqrt{E ({\rm GeV})}$ \\ 
\noalign{\vskip -2mm}&(em GeV)&(em GeV)&(\%)&(\%)\\ \hline \noalign{\vskip -1.5mm}
20 GeV $\pi^+$&8.00&20.5&6.61&29.5  \\  \noalign{\vskip -1mm}
20 GeV $p$&6.76&20.2&6.48&29.0  \\
40 GeV $\pi^+$&21.7&41.3&4.49&28.4  \\  \noalign{\vskip -1.5mm}
40 GeV $p$&18.5&40.7&4.38&27.6  \\
%\vskip 1mm
60 GeV $\pi^-$&38.5&61.0&3.90&30.2  \\
%\vskip 1mm
80 GeV $\pi^+$&51.5&80.7&3.22&28.8  \\  \noalign{\vskip -1.5mm}
80 GeV $p$&47.1&80.4&3.34&29.9  \\
%\vskip 1mm
125 GeV $\pi^+$&84.8&127&2.63&29.4  \\  \noalign{\vskip -1.5mm}
125 GeV $p$&77.8&126.5&2.85 &31.9  \\ \hline
\end{tabular}
\label{rotateres}
\end{table}

The results are summarized in Table \ref{rotateres}, which lists for each type of particle the average value of the measured \v{C}erenkov 
signals, the average signal after application of the dual-readout rotation method, the fractional energy resolution ($\sigma/E$) and the 
fractional energy resolution multiplied with $\sqrt{E}$. All signal values are expressed in em GeV, \ie the energy scale derived from the calibration with electron showers. 
\begin{figure}[htbp]
\epsfysize=7cm
\centerline{\epsffile{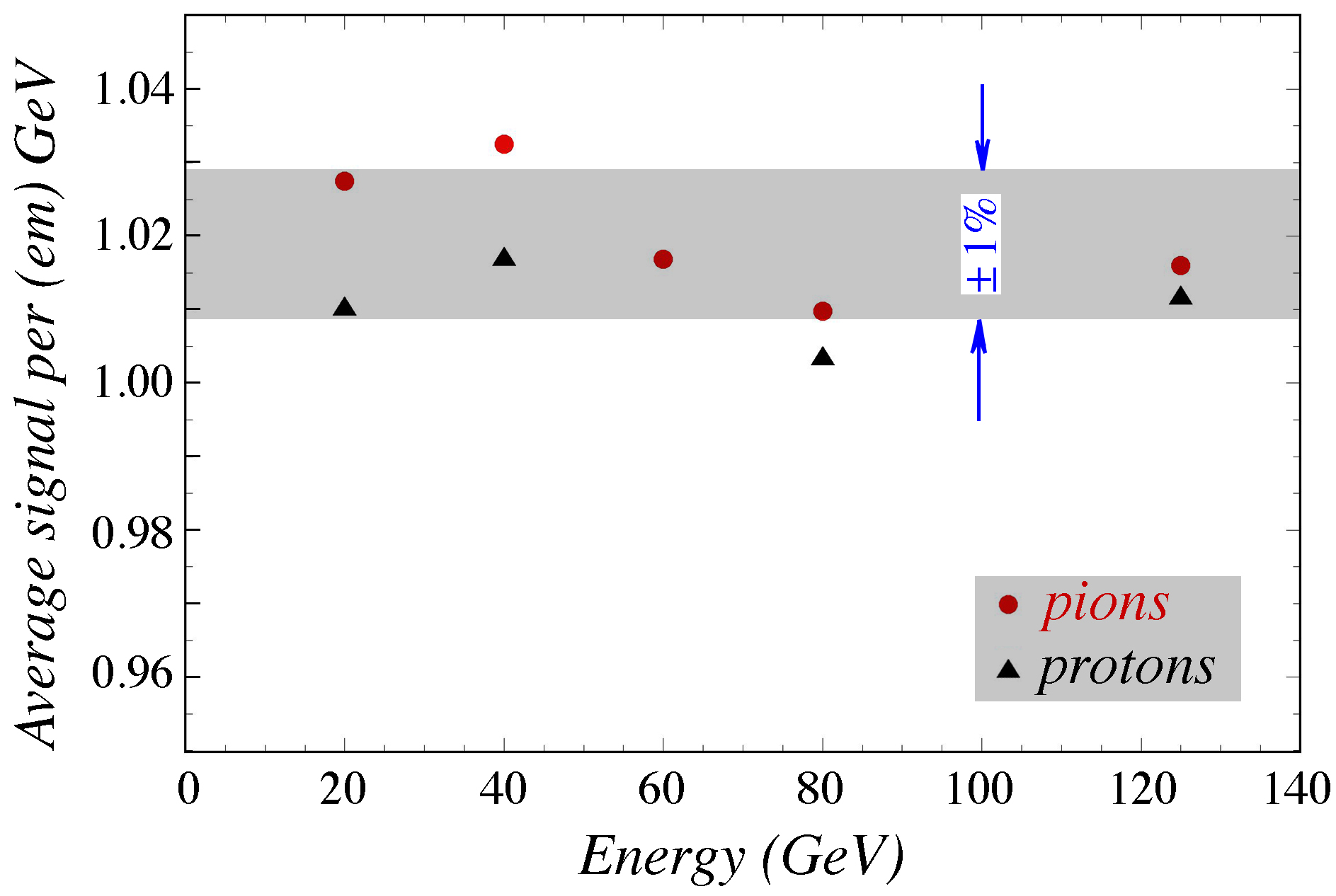}}
\caption{\footnotesize{The calorimeter response, \ie the average signal for protons and pions per GeV, as a function of energy.
The vertical scale is normalized to the electron response. }}
\label{linrot}
\end{figure}
\vskip 10mm
These results exhibit some very important features:
\begin{itemize}
\item The calorimeter is very linear, both for pion and for proton detection. The beam energy is correctly reconstructed at all energies within a few percent, using the energy scale for electrons, which were used to calibrate the signals. Figure \ref{linrot} shows the calorimeter response to protons and pions, \ie the average signal per unit deposited energy, as a function of energy. Variations of $\pm 1\%$ about the average value are indicated by the shaded band. The vertical scale is normalized to the electron response. The hadron signals are thus a 
few percent larger than those for em showers of the same energy. 
\item The reconstructed signal distributions are very narrow, narrower  than those reported by any other detector we know of.
\begin{figure}[b!]
\epsfysize=5.3cm
\centerline{\epsffile{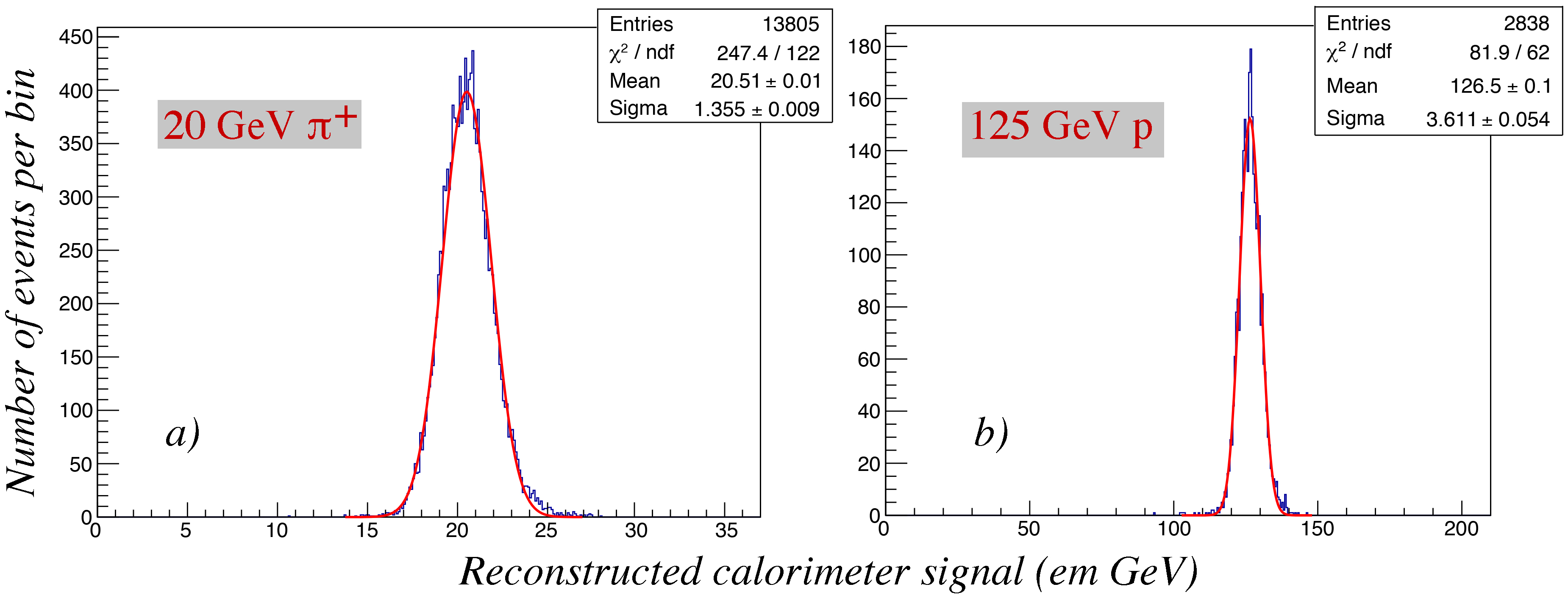}}
\caption{\footnotesize{Signal distributions for 20 GeV $\pi^+$ ($a$) and 125 GeV protons ($b$) obtained with the rotation method described in the text. The energy scale is set by electrons showering in this detector.}}
\label{20pi125p}
\end{figure}
\item The reconstructed signal distributions are very well described by Gaussian functions. This is illustrated in Figure \ref{20pi125p}, which shows signal distributions for hadrons at the low and high end of the spectrum of particles studied here. The normalized $\chi^2$ values
varied between 1.02 and 2.27 for all particles listed in Table \ref{rotateres}.
\item The fractional width of the reconstructed signal distribution also scales very well as expected for an energy resolution dominated by Poissonian fluctuations. Over the full energy range of 20 - 125 GeV we find: $\sigma/E = (30 \pm 2\%)/\sqrt{E}$. This result is represented by the straight line in Figure  \ref{resrot}, which shows the experimental data points, separately for protons and pions, as a function of the beam energy.
\begin{figure}[htbp]
\epsfysize=7.5cm
\centerline{\epsffile{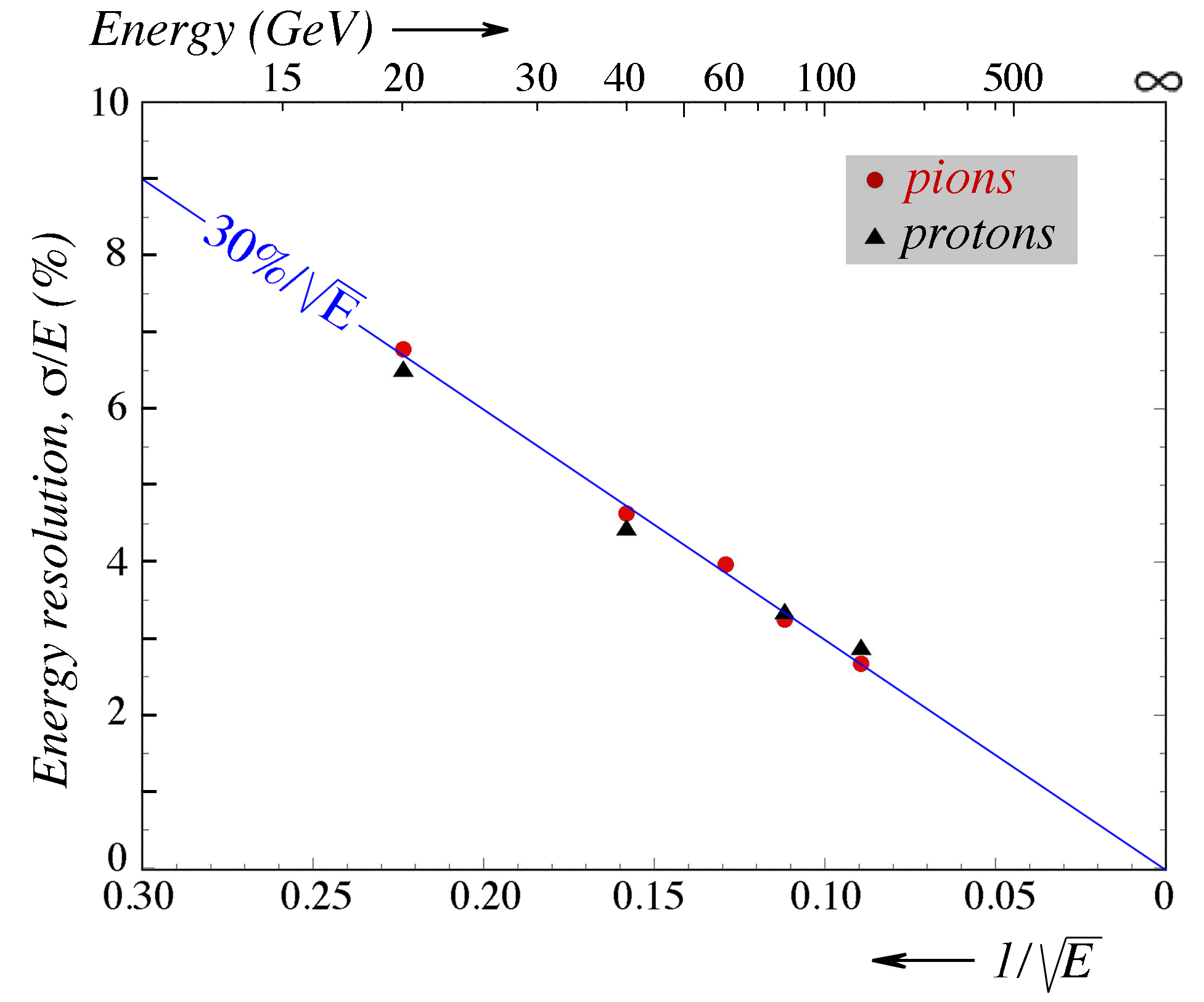}}
\caption{\footnotesize{The fractional width of the signal distribution, $\sigma/E$, as a function of energy, for pions and protons in the 20 - 125 GeV energy range. The line represents $\sigma/E= 30\%/\sqrt{E}$.}}
\label{resrot}
\end{figure}
\end{itemize}

\subsection{The rotation method for multiparticle events}
\vskip -5mm

This method was used with the same rotation angle ($\theta = 30^\circ$) for multiparticle events, samples of which were available for beam energies of +40, +60, +100 and +125 GeV.
During these dedicated runs, the Interaction Target was installed in the beam line (see Figure \ref{layout}). Events were selected by requiring that the beam hadrons produced a signal compatible with a mip in the upstream PSD and a signal of at least 6 mip in the downstream scintillation counter. No distinction was made between protons and pions for this analysis. 
Otherwise, the conditions were identical to the ones used for the single-hadron analysis.
\begin{figure}[htbp]
\epsfysize=7cm
\centerline{\epsffile{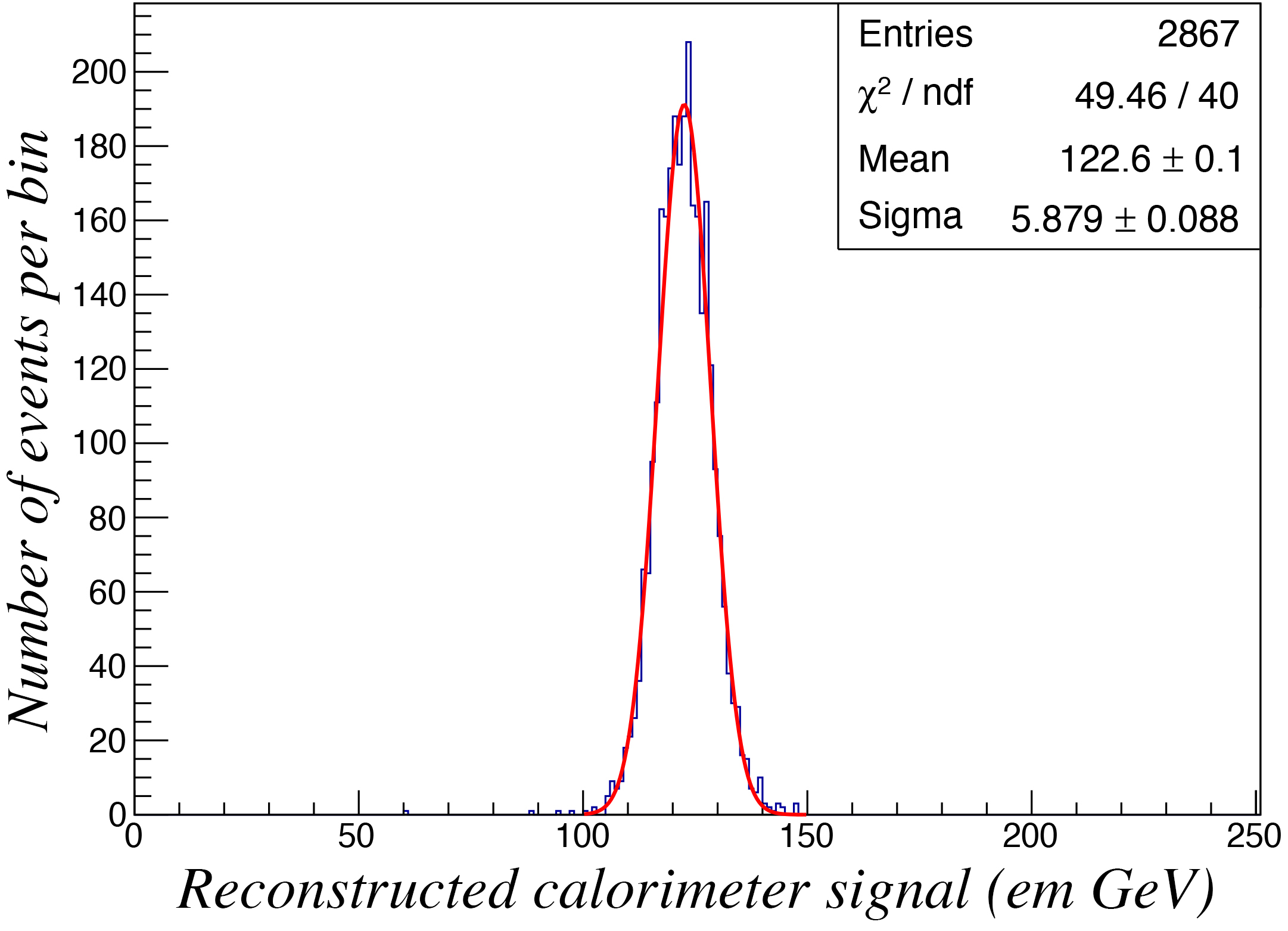}}
\caption{\footnotesize{Signal distribution for 125 GeV multiparticle events obtained with the rotation method described in the text. The energy scale is set by electrons showering in this detector.}}
\label{125jet}
\end{figure}

\begin{figure}[htbp]
\epsfysize=7.5cm
\centerline{\epsffile{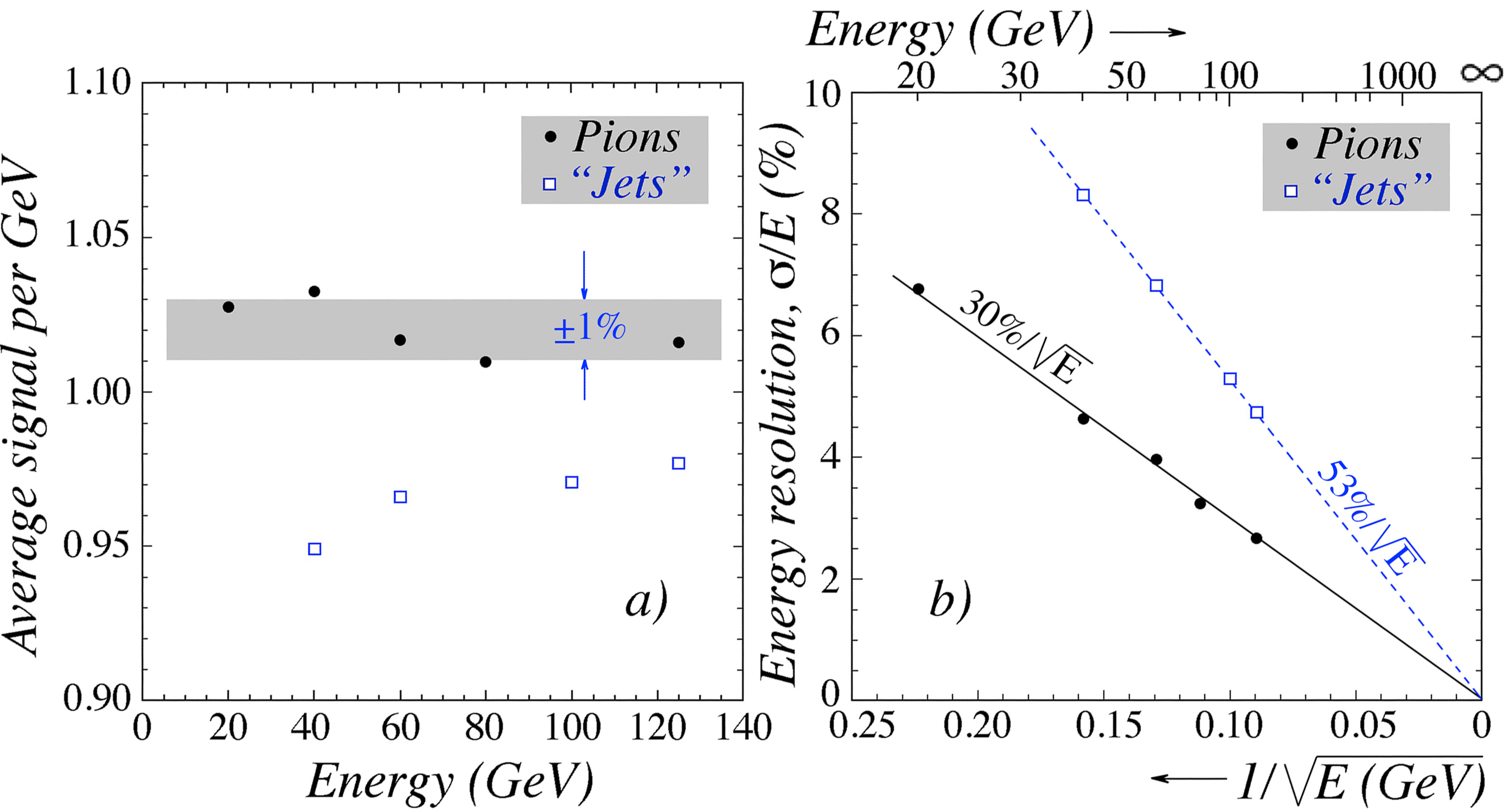}}
\caption{\footnotesize{The average calorimeter signal per GeV ($a$) and the fractional width of the signal distribution ($b$) as a function of energy, for single pions and multiparticle events (``jets''). Results are given for the dual-readout calorimeter signals, obtained with the rotation method.}}
\label{jetresults}
\end{figure}

Figure \ref{125jet} shows an example of the signal distribution for 125 GeV multiparticle events obtained with the rotation method. This distribution shows similar features as those for single hadrons (Figure \ref{20pi125p}): A rather narrow distribution, centered at approximately the correct (energy) value, well described by a Gaussian function. However, there are also some differences, which become more obvious when we look at the results for all energies for which this analysis was carried out. These are listed in Table \ref{rotatejet}, and shown graphically in Figure \ref{jetresults}. 
\begin{table}[hbtp]
\caption{\footnotesize The reconstructed energy and the energy resolution for showers induced by pions and by multiparticle events (``jets''), measured with the rotation method. See text for details.}
\centering
\vskip 3mm
\begin{tabular} {|lcccc|} \hline
{\sl Particles}&$\langle$\v{C} signal$\rangle$&$\langle$Reconstructed energy$\rangle$&$\sigma/E$ (\%)&${\sigma/E}\cdot \sqrt{E ({\rm GeV})}$ \\ 
\noalign{\vskip -2mm}&(em GeV)&(em GeV)&(\%)&(\%)\\ \hline \noalign{\vskip -1.5mm}
40 GeV $\pi^+$&21.7&41.3&4.49&28.4  \\  \noalign{\vskip -1.5mm}
40 GeV ``jets''&~~~~~~14.7&~~~~~~37.9&~~~~~~8.32&~~~~~~52.6  \\
%\vskip 1mm
60 GeV $\pi^-$&38.5&61.0&3.90&30.2  \\ \noalign{\vskip -1.5mm}
60 GeV ``jets'' &~~~~~~27.6&~~~~~~58.0&~~~~~~6.83&~~~~~~52.9  \\
%\vskip 1mm
%100 GeV $\pi^+$&51.5&80.7&3.22&28.8  \\  \noalign{\vskip -1.5mm}
100 GeV ``jets''&~~~~~~54.9&~~~~~~97.1&~~~~~~5.30&~~~~~~52.9  \\
%\vskip 1mm
125 GeV $\pi^+$&84.8&127&2.63&29.4  \\  \noalign{\vskip -1.5mm}
125 GeV ``jets''&~~~~~~69.0&~~~~~~122.6&~~~~~~4.79&~~~~~~53.6 \\ \hline
\end{tabular}
\label{rotatejet}
\end{table}

It turns out that the multiparticle signal distributions are clearly wider than those for single hadrons. However, in both cases, the fractional width scales with $E^{-1/2}$, without any significant deviations: $53\%/\sqrt{E}$ for ``jets'', \vs $30\%/\sqrt{E}$ (Figure \ref{jetresults}b). This indicates that only stochastic fluctuations contribute to this width.
The reconstructed energies are also somewhat lower in the case of the multiparticle events, more so at low energy (Figure \ref{jetresults}a). Very substantial differences are observed in the size of the \v{C}erenkov component, which is on average considerably smaller for the multiparticle events. 

These features can be understood by realizing that the primary interaction of the beam particles took place at a distance of about 75 cm upstream of the calorimeter.
Low-energy secondaries produced in these interactions may have traveled at such large angles with the beam line that they physically missed the calorimeter, as well as the leakage counters surrounding the calorimeter. The effect of that is larger when the energy of the incoming beam particle is smaller. The increased side leakage is probably also the main factor responsible for the increased width of the signal distribution.
The difference in the strength of the \v{C}erenkov component most likely reflects the fact that the average energy fraction carried  by the em component in hadronic showers increases with energy. Therefore, if the energy of the incoming beam particle is split between at least six secondaries (our trigger condition for multiparticle events), the total em energy fraction is likely to be smaller than when the beam particle enters the calorimeter
and deposits its entire energy there in the form of a single hadronic shower.  

\subsection{Discussion}
\vskip -5mm

Notice that we have {\sl not} used any knowledge about the energy of the beam particles in the rotation procedure described in the previous subsections. The coordinates of the rotation center were chosen on the basis of the {\sl equality of the hadronic \v{C}erenkov and scintillation
signals}. This implies that the hadronic response at that point must be equal to that for electrons, which was used to set the energy scale
for both types of calorimeter signals. 
The described method has thus allowed us to measure the energy of the beam particles with great precision. The average beam energy has been correctly reproduced within a few percent  for all energies studied, the fractional width of the signal distribution scaled with $E^{-1/2}$ and, most interestingly, the dual-readout signal distributions were found to be essentially identical for protons and pions, despite the substantial differences between the signal distributions for these particles measured in the scintillation or \v{C}erenkov channels. The latter aspect is a unique feature of dual-readout calorimetry.
No other calorimeter we know of is capable of this. ATLAS has reported significant differences between the signal distributions of protons and pions \cite{Adr10}, but their ``offline compensation'' methods required prior knowledge of the particle type to eliminate these differences. 

Yet, while we have managed to obtain very narrow signal distributions for the beam particles using only the calorimeter information, we don't think it is correct to interpret the relative width of these distributions as a measure for the precision with which the energy of an arbitrary particle absorbed in this calorimeter may be determined. The determination of the coordinates of the rotation point, and thus the energy scale of the signals, relied on the availability of an ensemble of events obtained for particles of the same energy. In practice, however, one is only dealing with {\sl one} event, of unknown energy, and the described procedure can thus not be used in that case.
\begin{figure}[htbp]
\epsfysize=6.5cm
\centerline{\epsffile{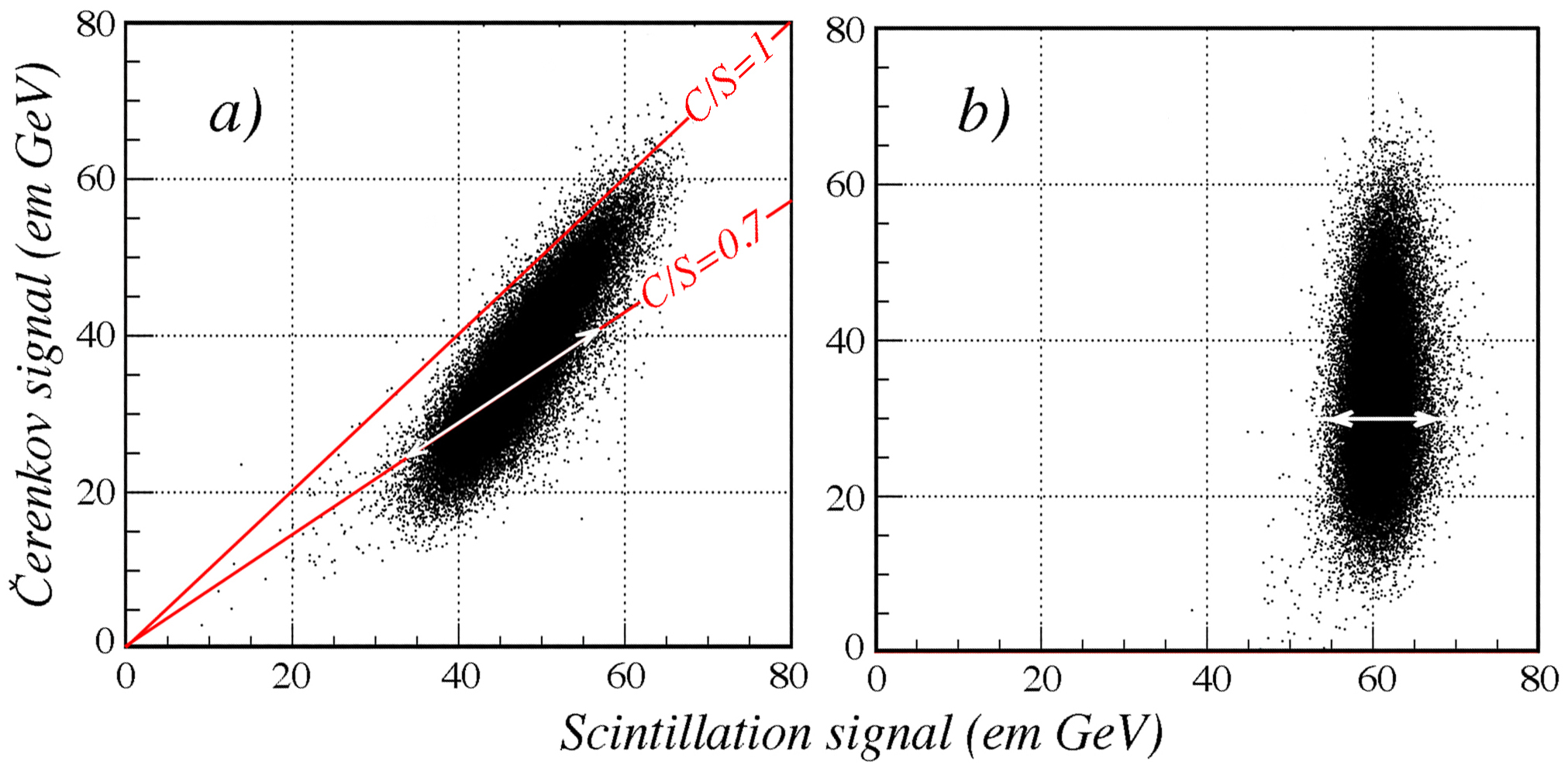}}
\caption{\footnotesize{Scatter plots of the \v{C}erenkov \vs the scintillation signals from showers induced by mono-energetic hadrons ($a$). 
The arrow indicates the precision with which the em shower fraction, and thus the energy, of an individual particle can be determined on the basis of the measured ratio of the \v{C}erenkov and scintillation signals, 0.7 in this example. The rotation procedure for an ensemble of 
mono-energetic pions leads to the scatter plot shown in diagram $b$.
The precision of the measurement of the width of that distribution is indicated by a white arrow as well.}}
\label{dr2method}
\end{figure}

The DREAM Collaboration has developed a procedure to determine the energy of an unknown particle showering in the dual-readout calorimeter that is {\sl not} affected by this problem. In this procedure, described in Section 3.1, the em shower fraction ($f_{\rm em}$) of the hadronic shower is derived from the ratio of the \v{C}erenkov and scintillation signals. Using the known $e/h$ values of the two calorimeter structures, the measured signals can then be converted to the em energy scale ($f_{\rm em} = 1$). 
The energy resolutions obtained with this method are worse than the ones given in this section, although it should be mentioned that they are 
dominated by incomplete shower containment and the associated leakage fluctuations, and are likely to improve considerably for detectors that are sufficiently large \cite{geant}. However, the same is probably true for the measurements of which the results are shown in Figure \ref{DRrotation}.
Figure \ref{dr2method} graphically illustrates the difference between the values of the energy resolution obtained with the two methods discussed here. The precision of the energy measurement is 
represented by the arrows in the two diagrams.
\vskip 2mm

The message we want to convey in this section is that one should not confuse the precision of the energy determination of a given event based on calorimeter signals alone with the width of a signal distribution obtained in a testbeam, since the latter is
typically based on additional information that is not available in practice. In the example described above, this additional information derived from the fact that a large number of events generated by particles of the same energy were available. In other cases, additional information may be derived from knowledge of the particle energy. This is especially true for calorimeters whose energy scale depends on ``offline compensation,'' or other techniques intended to minimize the total width of the signal distribution from a detector system consisting of several longitudinal segments. Such techniques rely on calibration constants whose values depend on the energy, on the type of showering particle, and sometimes also on the ratios of the signals from the different calorimeter sections.

\section{Conclusions}
\vskip -5mm

We have studied the hadronic performance of a lead-based dual-readout fiber calorimeter with beams of pions and protons of different energies, and with multiparticle events created by upstream interactions of these beam particles in a dedicated target. The assessment of the performance characteristics, and thus of the potential possibilities of this type of detector, was limited by the fact that the 
calorimeter was too small and used lead as absorber material. As has been pointed out elsewhere \cite{Wig13}, a lower-$Z$ absorber material such as copper would be much more suitable for this type of detector. However, we have not yet managed to identify a low-cost technique for mass production of the complicated absorber structure out of copper. On the other hand, lead could be extruded into the desired shape.

We have demonstrated that the hadronic energy resolution of the tested calorimeter was dominated by fluctuations in lateral shower leakage. We have tried to mitigate these effects with a crude and rather non-hermetic system of leakage counters surrounding the calorimeter. This certainly improved the energy resolution significantly, but not nearly enough to eliminate the leakage effects. The effects of leakage on the energy resolution became clear by selecting events in which no measurable leakage occurred. For these events, the measured resolution was comparable to that of the best compensating calorimeters ever built.

Similar performance was achieved with an analysis method in which we made use of the availability of an ensemble of events caused by particles of the same energy. The availability of two signals that provided complementary information about the showers made it possible to determine the energy of the particles, independent of any additional information. A simple rotation procedure then led to signal distributions with all the characteristics of an ideal calorimeter: Signal linearity, Gaussian response functions with a very narrow width that scaled with $E^{-1/2}$, and the same response for pions and protons, whose responses differed substantially when measured in the scintillation or \v{C}erenkov channels. 

With the exception of the energy resolution, similarly good performance was obtained with the standard dual-readout method, which can be applied for individual events. The width of the signal distribution which, as explained above, was dominated by lateral leakage fluctuations, was also in this case measured to be completely determined by stochastic fluctuations, as evidenced by the $E^{-1/2}$ scaling.

%\newpage
\section*{Acknowledgments}
\vskip -5mm

We thank CERN for making good particle beams available to our experiments in the H8 beam. 
In particular, we also thank the technicians who are responsible for the construction and installation of the calorimeter: Freddi Angelo, Domenico Calabr\`o, Claudio Scagliotti and Filippo Vercellati.
This study was carried out with financial support of the United States
Department of Energy, under contract DE-FG02-12ER41783, of Italy's Istituto Nazionale di Fisica Nucleare and Ministero dell'Istruzione, dell' Universit\`a e della Ricerca, and of the Basic Science Research Program of the National Research Foundation of Korea (NRF), funded by the Ministry of  Science, ICT \& Future Planning under contract 2015R1C1A1A02036477. 

\bibliographystyle{unsrt}

\begin{thebibliography}{99.}

\bibitem{Akc97} N.~Akchurin \etal, \NIM {\bf A399} (1997) 202.

\bibitem{RD52web} 
All publications and other results obtained in the context of this project can be found at the RD52 website:\hb
{\it http://highenergy.phys.ttu.edu/dream}\hb
{\it http://dream.knu.ac.kr}

%\bibitem{TeVpaper} R. Wigmans, New Journal of Physics {\bf 10} (2008) 025003.

%\bibitem{Springer} R. Wigmans, in {\sl Handbook of Particle Detection and Imaging}, eds. C. Grupen and I. Buvat, vol. 1, 497-517, Springer Verlag (2011).

\bibitem {RD52_em} N.~Akchurin \etal, \NIM {\bf A735} (2014) 130.

\bibitem{smalltheta} A.~Cardini \etal, \NIM {\bf A808} (2016) 41.

\bibitem{PID} N.~Akchurin \etal, \NIM {\bf A735} (2014) 120.

\bibitem{Akc05a} N. Akchurin \etal, \NIM  {\bf A537} (2005) 537.

\bibitem{deg07} D.E. Groom, \NIM {\bf 572} (2007) 633.

\bibitem{Aco90} D.~Acosta \etal, \NIM {\bf A305} (1991), 55.

\bibitem{Hartjes} F.G. Hartjes and R. Wigmans, \NIM {\bf A277} (1989), 379.

\bibitem{geant} N.~Akchurin \etal, \NIM {\bf A762} (2014) 100.

\bibitem {Akc14} N.~Akchurin \etal, \NIM {\bf A735} (2014) 120.

\bibitem{Akc98}  N.~Akchurin \etal, \NIM {\bf A408} (1998) 380.

\bibitem{Wig00}
R.~Wigmans, {\em Calorimetry, Energy Measurement in Particle Physics}, International Series of
Monographs on Physics, Vol. 107, Oxford University Press (2000).

\bibitem{PDG16} C.~Patrignani \etal ~(Particle Data Group),  {\em Chin. Phys.} {\bf C40}, 100001 (2016), Section 34.9.2.

\bibitem{Adr10} P.~Adragna \etal, \NIM {\bf A615} (2010) 158.

\bibitem{Wig13}
R.~Wigmans, \NIM {\bf A713} (2013) 43.

%\bibitem{DREAMmu} N.~Akchurin \etal, \NIM {\bf 533} (2004) 305.

%\bibitem{Elba15} R.~Wigmans, {\sl New Results from the RD52 Project}, Proc. of the 13th Pisa Meeting on ``Frontier Physics for Frontier Detectors", May 24-30 2015, La Biodola, Elba (Italy), submitted to \NIM

%\bibitem{Dre90} G.~Drews \etal, \NIM {\bf A290} (1990) 335.

%\bibitem{Aco91a} D.~Acosta \etal, \NIM {\bf A308} (1991) 481.

%\bibitem{Fabjan} C.W.~Fabjan \etal, \NIM {\bf 141} (1977) 61.

%\bibitem{DREAMhad} N.~Akchurin \etal, \NIM {\bf A537} (2005) 537.

%\bibitem{Aco91a} D.~Acosta \etal, \NIM {\bf A302} (1991) 36.

%\bibitem{olga} M.~Albrow \etal, \NIM {\bf A487} (2002) 381.

%\bibitem{calor06} R.~Wigmans, {\sl On the calibration of segmented calorimeter systems}, Proc. XIIth Int. Conf. on Calorimetry in High Energy Physics, Chicago 2006, AIP Conf. Proc. {\bf 867} (2006) 90.

%\bibitem{multiv} M.~Backes \etal, {\sl TMVA 4, Toolkit for Multivariate Data Analysis with ROOT}, arXiv:physics/0703039.

%\bibitem{Hubbell} J.H.~Hubbell, {\sl Photon cross sections, attenuation coefficients and energy absorption coefficients from 10 keV to 100 GeV}, report NSRDS-NBS 29 (1969).

%\bibitem{Ask} G.A.~ Askar'yan, Sov. Phys. JETP {\bf 14} (1961) 441; {\bf 21} (1965) 658.

%\bibitem{Ask1} D. Saltzberg \etal, \prl {\bf 86} (2001) 2802.

%\bibitem{Ask2} P.W.~Gorham \etal, \prl {\bf 99} (2007) 171101.

%\bibitem{GEANT4} S.~Agostinelli \etal, \NIM {\bf A506} (2003) 250.

%\bibitem{NJP08} R.~Wigmans, New Journal of Physics {\bf 10} (2008) 025003.

%\bibitem{Cproof} N.~Akchurin \etal, \NIM {\bf A582} (2007) 474.

%\bibitem{temp} N.~Akchurin \etal, \NIM {\bf A593} (2008) 530.

%\bibitem{Sep08} N.~Akchurin \etal, \NIM {\bf A595} (2008) 359.

%\bibitem{Kleb99} E.O.~Klebanski\u{\i} \etal, Phys. of the Solid State {\bf 41} (1999) 913.

%\bibitem{Yan05} Z.~Yang, Q.~Zhang and Z.~Jiang, J. Phys. D: Appl. Phys. {\bf 38} (2005) 1461.

%\bibitem{newcrystals} N.~Akchurin \etal, \NIM {\bf A604} (2009) 512.

%\bibitem{DREAMem} N.~Akchurin \etal, \NIM {\bf A536} (2005) 29.

%\bibitem{Shi05} H.~Shimizu \etal, \NIM {\bf A550} (2005) 258.

%\bibitem{nikl} M.~Nikl \etal, Journal of Applied Physics {\bf 104} (2008) 1.
%\bibitem{niklMo} M.~Nikl \etal, Journal of Applied Physics {\bf 91} (2002) 2791.

%\bibitem{ECALPbWO} N.~Akchurin \etal, \NIM {\bf A584} (2008) 273.

\end{thebibliography}

\end{document}